\documentclass[preprint,12pt]{elsarticle}

\usepackage{color,array}
\usepackage{ulem}
\usepackage{graphicx}
\usepackage{color}
\usepackage{stmaryrd}
\usepackage{comment}
\usepackage{graphicx}
\usepackage{verbatim}
\usepackage{diagbox}
\usepackage{indentfirst}
\usepackage{makecell}
\usepackage{subfigure}
\usepackage{algorithm}
\usepackage{algorithmicx}
\usepackage[noend]{algpseudocode}
\usepackage{soul,color,xcolor}
\usepackage{pifont}
\usepackage{amsmath, amsthm, amssymb, paralist}
\usepackage{url}
\usepackage[colorlinks,breaklinks,urlcolor=blue,linkcolor=blue,citecolor=blue]{hyperref}
\usepackage{xurl}
\usepackage{breakurl}
\soulregister\cite7
\soulregister\ref7
\soulregister\rsubsec7
\newcommand{\req}[1]{\eqref{#1}}
\newtheorem{mydef}{Definition}
\newtheorem{myprop}{Proposition}
\newtheorem{mypro}{Problem}

\newtheorem{mythm}{Theorem}

\newtheorem{remark}{Remark}
\newcommand{\qedwhite}{\hfill \ensuremath{\Box}}
\newcommand{\rsec}[1]{Section\,\ref{#1}}
\newcommand{\rsubsec}[2]{Section\,\ref{#1}-\ref{#2}}
\newcommand{\rdef}[1]{Definition\,\ref{#1}} 
\newcommand{\rprop}[1]{Proposition\,\ref{#1}} 
\newcommand{\rpro}[1]{Problem\,\ref{#1}}
\newcommand{\rlem}[1]{Lemma\,\ref{#1}}
\newcommand{\ralg}[1]{Algorithm\,\ref{#1}}
\newcommand{\rthm}[1]{Theorem\,\ref{#1}}
\newcommand{\rtab}[1]{Table\,\ref{#1}}
\newcommand{\rfig}[1]{Fig.\;\ref{#1}}
\newcommand{\rrem}[1]{Remark\;\ref{#1}}

\usepackage{amsmath}
\usepackage{cases}
\usepackage{amsfonts, amssymb}

\usepackage{amssymb}

\journal{Nonlinear Analysis Hybrid Systems}

\begin{document}

\begin{frontmatter}



\title{Synthesis of Event-triggered Controllers for SIRS Epidemic Models}


\author[inst1]{Lichen Ding}

\affiliation[inst1]{organization={Graduate School of Engineering, Osaka University, Japan},}

\author[inst1]{Kazumune Hashimoto}
\author[inst1]{Shigemasa Takai}

\begin{abstract}
In this paper, we investigate the problem of mitigating epidemics by applying an event-triggered control strategy. 
\textcolor{black}{We consider a susceptible-infected-removed-susceptible (SIRS) model, which builds upon the foundational SIR model by accounting for reinfection cases.} The event-triggered control strategy is formulated based on the condition in which the control input (e.g., the level of public measures) is updated \textit{only when} {the fraction of the infected subjects} {changes} over a designed threshold. 
To synthesize the event-triggered controller, we leverage the notion of a symbolic model, which represents an abstracted expression of the transition system associated with the SIRS model under the event-triggered control strategy. 
\textcolor{black}{Then, by employing safety and reachability games, two event-triggered controllers are synthesized to ensure a desired specification, which is more sophisticated than the one given in previous works.} 
The effectiveness of the proposed approach is illustrated through numerical simulations.
\end{abstract}


\end{frontmatter}


\section{Introduction}\label{introduction}

\textcolor{black}{For many years, infectious diseases have caused numerous epidemics, leading to millions of fatalities and infections and socioeconomic consequences. Research in mathematical modeling of these diseases, mainly through the use of ordinary differential equations (ODEs), has been of significant importance over the past decades. These mathematical models assist in not only predicting disease propagation over time but also facilitating the analysis and synthesis of prevention strategies, such as vaccination and effective pharmaceutical treatments and non-pharmaceutical treatments (e.g., social distance, lockdown).} 
    To date, a wide range of mathematical models have been proposed, with the foundation laid by \cite{kermack}, and expanding to include numerous related or more sophisticated variants, see \cite{brauer2012mathematical} and references therein. Epidemic models have been utilized to systematically design treatment and vaccination strategies and mitigate epidemics based on the usage of control theory. 
In recent years, various control strategies (with the usage of control theory) have been proposed to explore an effective way of {mitigating epidemics}. \cite{xiao2012sliding} proposed a sliding mode control strategy to develop a systematic way of designing simple implementable controls that drive the dynamical system associated with the SIR model to a desired globally stable equilibrium. \cite{chen2014global} proposed a feedback control strategy that by constructing a Lyapunov function with suitable values of feedback control variables, the global stability of the disease-free equilibrium and the endemic equilibrium of the system associated with the SI model is investigated. 
\cite{casella2020can} shows that, by analyzing a variant of an SEIR model incorporated with time-delay, 
suppression strategies are shown to be effective if public measures are oppressive enough and enacted early on. 
\textcolor{black}{In addition, some approaches aim at designing vaccination or treatment strategies based on \textit{optimal control}, see, e.g., \cite{gaff2009,kar2011stability,yusuf2012optimal,tsay2020modeling,kohler2021robust}. For instance, \cite{gaff2009} investigated an optimal vaccination and treatment strategy for a SIR model by minimizing the sum of the number of infected people and the cost of vaccination and treatment and then deriving a necessary condition of optimality (Pontryagin maximum principle).} 
Most of the aforecited works on designing control strategies for epidemic models assume that the control inputs are updated continuously or per a fixed time period (e.g., 7 days), which means that the level of public measures is \textit{assumed} to be updated continuously or periodically. 
In view of the measure regulation in actual situations, such as those of COVID-19, however, it may be more natural and appropriate to update control inputs according to the increase/decrease of the infected people, e.g., the lockdown is carried out when the number of the infected people becomes larger to some extent, and we gradually loosen the measures according to how the number of the infected people decreases. Event-triggered control framework is suited for such a situation, as it is known to update the control input \textit{only when} the observed state exceeds a certain (prescribed) threshold while not compromising the control performance, see, e.g., \cite{heemels2012a} and reference therein. 
Although designing the event-triggered control framework for epidemic models seems to be more appropriate, only a very few works on these related topics are in the literature, see, e.g., \cite{hashimoto2020event}. 
In \cite{hashimoto2020event}, the authors investigated the event-triggered control design for SIS epidemic processes. The proposed event-triggered strategy updated the control input only when {the fraction of infected subjects} {changes} over a given threshold, which was designed by solving the geometric program. In addition, they showed that the resulting strategy guarantees ultimate boundedness property, in the sense that the fraction of infected subjects eventually decreases within the prescribed thresholds in a finite time.

\textcolor{black}{In this paper, we investigate the problem of designing novel event-triggered controllers for epidemic models, aiming to contain the epidemics, reduce the frequency of update of control inputs, and minimize the total control effort.} 
Here we focus on the susceptible-infected-removed-susceptible (SIRS) model \cite{brauer2012mathematical}, which is an extension of the fundamental SIR model by taking reinfection cases into consideration, and design not only the control input but also when it needs to be updated according to the increase or decrease of the fraction of infected subjects (e.g., raising the level of the measure if the fraction of infected subjects increases by $1\%$, or lowering the level of the measure if the fraction of infected subjects decreases by $2\%$). 
To specifically capture the dynamical behaviors of the epidemic spreading, a more sophisticated mathematical model might be required, such as the SIDARTHE model (see, e.g., \cite{giordano2020modelling}) for the current COVID-19 pandemic. Nevertheless, the SIRS model has still been utilized as a useful tool to capture essential behaviors of the epidemic spreading even for COVID-19 (see, e.g., \cite{SALIMIPOUR2023,Koichiro}). 
Hence, we argue that it is worth investigating 
\textit{when} and \textit{how} public measures should be taken to contain the spread of infection based on the SIRS model. 
In addition, this paper can be viewed as the first step to synthesizing the \textit{event-triggered control policy} for epidemic models with \textit{complex specifications} (as detailed below). 
\textcolor{black}{The main novelty of our approach with respect to the related previous works, in particular to \cite{hashimoto2020event}, is twofold. First, our approach allows us to design an event-triggered control strategy with reachability and {safety} assurance with respect to the fractions of susceptible and infected subjects, which are more sophisticated specifications than those in the previous works (e.g., \cite{hashimoto2020event}). Specifically, we can show in this paper that the fraction of susceptible subjects consistently remains above a specified threshold, while the fraction of infected subjects is {always} below a given {upper limit} and eventually falls below a specific level. 
This is crucially different from previous works such as \cite{hashimoto2020event}, where only an ultimate boundedness property of the infected subjects (i.e., the fraction of infected subjects decreases eventually within the prescribed thresholds) or stability is analyzed. 
Second, our approach allows us to deal with the \textit{discrete} nature of the control inputs. In general, public measures (control inputs) to {contain epidemics} are discrete (``no interventions", ``lockdown", \textit{etc}.), and so with the control inputs being matched up with different levels of measure, we are able to determine which measures should be taken when updating them. 
Thus, such discrete control inputs are more applicable and suitable compared to those designed in previous works, which take values within a \textit{continuous} range (e.g., $u(t) \in [0, 1]$).} 

In this paper, the aforementioned novelties are achieved by employing a \textit{symbolic model} (see, e.g., \cite{tabuada2009verification}). 
In general, the symbolic model represents an abstracted expression of the control system, where each state of the symbolic model corresponds to an aggregate of the states of the control system. The utilization of the symbolic model allows us to synthesize controllers that guarantee the specifications considered in this paper by employing algorithmic techniques, such as reachability and safety games~\cite{tabuada2009verification}. 
To employ the symbolic approach for synthesizing event-triggered controllers, we first propose a way of constructing a transition system that incorporates {an event-triggered strategy into the SIRS model}, which we will call an ``event-triggered transition system" (for details, see \rsec{event-triggered_transition_system}). In essence, this transition system captures the fact that the control input is updated only when the fraction of infected subjects {changes} over a given threshold. Then, we propose a way of constructing the corresponding symbolic model, which represents an abstraction of the event-triggered transition system. Specifically, the symbolic model is constructed to guarantee the existence of an approximate alternating simulation relation (for details, see \rsec{symbolic approach}). 
\textcolor{black}{Then, event-triggered control policies are synthesized by employing a \textit{reachability game}, which is an algorithmic technique to design a policy for a finite transition system such that the state trajectory reaches a prescribed set in finite time, and a \textit{safety game}, which is also an algorithmic technique to design a policy such that the state trajectory remains in a given set for all times.} 
Hence, it allows us to ensure that the fraction of susceptible subjects is always greater than or equal to the desired lower limit and the fraction of the infected subjects is always below the given upper limit. 

Summarizing, the main contribution of this paper is to propose a \textit{symbolic} approach to synthesizing a novel event-triggered controller for SIRS epidemic models. In particular, we provide a way of constructing a symbolic model to capture an approximate behavior of the SIRS model under the event-triggered strategy. 
Then, event-triggered controllers are synthesized by employing a safety game to fulfill the desired control objectives, which are more sophisticated than those in the previous works. Finally, the effectiveness of the proposed approach has been validated through numerical simulations. Here, we made use of some infection data in Tokyo to determine some candidates of the control inputs and then employed our approach accordingly (for details, see \rsec{numerical simulation}).

This paper is organized as follows. In \rsec{problem formulation}, we introduce the SIRS model and explain the event-triggered control strategy. In \rsec{solution approach}, a symbolic model is constructed based on the notion of an approximate alternating simulation relation {(ASR)}, and event-triggered policies are synthesized. 
In \rsec{numerical simulation}, simulation results are given to illustrate the effectiveness of the proposed approach. Conclusions and future
works are given in \rsec{conclusion and future work}.\\

\noindent
\textit{(Notation):} Given $x=[x_1, \ldots, x_n]^\top,y = [y_1, \ldots, y_n]^\top\in \mathbb{R}^n$, we denote $\preceq$ the component-wise vector order, i.e. $x\preceq y$ if and only if $x_i\leq y_i$ for all $i\in \left\{1,...,n\right\}$. Given $x, y \in \mathbb{R}^n$ with $x\preceq y$, let $\llbracket x,y \rrbracket \subset \mathbb{R}^n$ be the hyper-rectangle defined by the endpoints $x$ (lower-left) and $y$ (upper-right), i.e., $ \llbracket x,y \rrbracket=\left\{z\in \mathbb{R}^n : x\preceq z\preceq y\right\}$. We denote by $\|\cdot\|$ the Euclidean norm. Given $x\in \mathbb{R}^n,X\subset \mathbb{R}^n $, we denote by ${\rm Nearest}_X(x)$ the closest points in $X$ to $x$, i.e., ${\rm Nearest}_X(x)={\rm argmin}_{x'\in X}\|x-x'\|$. Given $X \subset \mathbb{R}^2$ and $\eta_{x_1}, \eta_{x_2} >0$, let $[X]_{\eta_{x_1}, \eta_{x_2}} = \{x = [x_1, x_2]^\top \in X: x_1 = n \eta_{x_1}, x_2 = m \eta_{x_2}, n, m \in \mathbb{Z}\}$ denote the discretization of the set $X$ with respect to the discretization parameters $\eta_{x_1}, \eta_{x_2}$. 
\textcolor{black}{Given $x \in \mathbb{R}$ and $\varepsilon >0$, we denote by $\mathcal{B} ({x}, \varepsilon) \subseteq \mathbb{R}^n$ the ball set of radius $\varepsilon$ centered at $x$ with respect to the infinity norm, i.e., $\mathcal{B}({x}, \varepsilon) = \{{x}' \in \mathbb{R}^n: \| {x}- {x}' \|_{\infty} \le \varepsilon \}$.}
Given $X \subset \mathbb{R}^2$ and $\eta_{x_1}, \eta_{x_2} >0$, let $\mathrm{Int}_{\eta_{x_1}, \eta_{x_2}} (X)$ denote an $(\eta_{x_1}, \eta_{x_2})$-interior set of $X$, i.e., $\mathrm{Int}_{\eta_{x_1}, \eta_{x_2}} (X) = \{x = [x_1, x_2]^\top \in X: \llbracket \underline{x}, \bar{x} \rrbracket \subseteq X, \underline{x} = [x_1-\eta_{x_1}, x_2-\eta_{x_2}]^\top, \bar{x} =  [x_1+\eta_{x_1}, x_2 + \eta_{x_2}]^\top \}$. In addition, let $\mathrm{Out}_{\eta_{x_1}, \eta_{x_2}} (X)$ denote an $(\eta_{x_1}, \eta_{x_2})$-exterior set of $X$, i.e., $\mathrm{Out}_{\eta_{x_1}, \eta_{x_2}} (X) = \{x' = [x'_1, x'_2]^\top \in \mathbb{R}^2: \exists x = [x_1, x_2]^\top \in X, x' \in \llbracket \underline{x}, \bar{x} \rrbracket,\ \underline{x} = [x_1-\eta_{x_1}, x_2-\eta_{x_2}]^\top, \bar{x} =  [x_1+\eta_{x_1}, x_2 + \eta_{x_2}]^\top \}$. Given a set $X$, we denote by $2^X$ the power set of $X$, which represents the collection of all subsets of $X$.

\section{Problem Formulation}\label{problem formulation}

\subsection{System description}\label{system description}
In this paper, we consider the \textcolor{black}{SIRS model} \cite{brauer2012mathematical}, which is represented by the set of ordinary differential equations:
\begin{align}
&\dfrac{dS(t)}{dt}=-u(t)S(t)I(t) \textcolor{black}{+\xi R(t)}, \label{state_s}\\
&\dfrac{dI(t)}{dt}=u(t)S(t)I(t)-{\gamma}I(t), \label{state_i}\\
&\dfrac{dR(t)}{dt}={\gamma}I(t) \textcolor{black}{-\xi R(t)}.   \label{state_r} 
\end{align}

The state variables $S(t)$, $I(t)$, and $R(t)$ represent the fraction of Susceptible, Infected, and Removed subjects at the time instant $t \geq 0$, respectively. Note that the total fraction $S(t)+I(t)+R(t)$ is always constant, and, without loss of generality, it is assumed that $S(t)+I(t)+R(t)\equiv 1$ (for all $t \geq 0$). \textcolor{black}{Moreover, 
$[S(t_0), I(t_0)]^\top \in X_0$, where $X_0 = \{[S, I]^\top \in [0,1]^2 : S \in  [\underline{S}_0, \bar{S}_0] , I\in [\underline{I}_0, \bar{I}_0] $, $S+I \leq 1 \}$ for given $\underline{S}_0, \bar{S}_0, \underline{I}_0, \bar{I}_0 \in (0, 1)$, 
i.e., {the initial fraction of susceptible and infected subjects can start from any point} in $X_0$.} $u(t) >0$ is the time-varying parameter that accounts for the infection rate at the time instant $t \geq 0$, \textcolor{black}{and $\gamma, \xi >0$ are the rate of recovery and of immunity loss, respectively}. We assume that $\gamma, \xi$ are constant for all times. 
In addition, from \req{state_i} we have 
\begin{align}
&\dfrac{dI(t)}{dt}=({u}(t)S(t)-{\gamma})I(t). \label{state_i2}
\end{align}
Hence, whether $I(t)$ increases or decreases instantaneously at $t$ is determined by the sign of ${u}(t)S(t)-\gamma$, i.e., it increases at $t$ if ${u(t)S(t)}/{\gamma}>1$ and decreases if ${u(t)S(t)}/{\gamma} <1$. In general, the quantity of ${u}(t)S(t)/{\gamma}$ is called an \textit{effective reproduction number} \cite{brauer2012mathematical}.

\textcolor{black}{In this paper, we assume that the time-dependent parameter $u(t)$ is a function of the manipulated variable, or the \textit{control input}. This premise is sound as it implies the infection rate could be modulated based on the implementation of public measures such as lockdowns, facility closures, and so on. The control input, furthermore, is constrained such that $u(t) \in U$ for every $t \geq 0$. The set $U$ includes elements $u_0, \ldots, u_{m-1}$ in a descending order, i.e., $u_0 > u_1 > \cdots > u_{m-1}$. The intervention spectrum ranges from $u_0$, signifying the infection rate in the absence of any public measures, to $u_{m-1}$, representing the most stringent interventions implemented by the authorities.} 
For instance, previous research such as \cite{kobayashi2020predicting}, \cite{Koichiro} divide the COVID-19 outbreak in Tokyo, Japan, into several stages (no intervention, state of emergency, \textit{etc}.) and have computed estimated values of the infection rate for each stage (for details, see \rsec{numerical simulation}). 

\textcolor{black}{Since $R(t) = 1 - S(t) - I(t)$, we can substitute this equation into \req{state_s}, \req{state_i} (i.e., we can eliminate the variable of $R(t)$) and focus only on the dynamics of $S$ and $I$.} 
Thus, we express \req{state_s}, \req{state_i} compactly as 
\begin{align}
& \dfrac{dx(t)}{dt}=f(x(t),u(t)), \label{system_f}
\end{align}
where $x=[S, I]^\top \in X$ with $X=\left\{[S, I]^\top \in [0,1]^2 : S+I\le{1}\right\}$ and $x(t_0)=[S(t_0),I(t_0)]^\top \in X_0$. $f: X \times U \rightarrow \mathbb{R}^2$ is {the function that describes the dynamics of the fraction of susceptible and infected subjects} and is appropriately defined from \req{state_s} and \req{state_i}. 
Given $x \in X$ and $u \in U$, we denote by $\phi_{t}^{f}(x,u)\in X$ the state of the system \req{system_f} reached at time $t$ starting from $x$ ($x(t_0) = x$) by applying the constant control input $u(t) = u$, $t\geq 0$. In addition, let $\phi_{t,S}^{f}(\cdot)$ and $\phi_{t,I}^{f}(\cdot )$ denote the $S$-value and the $I$-value (i.e., the first and the second element) of $\phi_{t}^{f}(\cdot)$, respectively. 
Moreover, given $X' \subset X$ and $u \in U$, we denote by $R_{t}^{f}(X', u) \subset X$ the set of reachable states at time $t$ starting from any $x \in X'$ under the constant control input $u \in U$, i.e., $R_{t}^{f}(X',u)=\left\{\phi_{t}^{f}(x,u)\in X : x\in X'\right\}$. 
Similarly, given {$x \in X$} and the control input until $t$, {$u(t'), t' \in [0, t]$}, we denote by $\phi_{t}^{f}(x,\mathbf{u}(t))\in X$ with $\mathbf{u}(t) = \{u(t') : t' \in [0, t]\}$ the state of the system \req{system_f} reached at time $t$ starting from $x$ ($x(t_0) = x$) by applying $u(t'), t' \in [0, t]$. 
In addition, given $X' \subset X$ and $U' \subset U$, we denote by $R_{t}^{f}(X', U') \subset X$ the set of reachable states at time $t$ starting from any $x \in X'$ under the realization of all possible control inputs $u(t') \in U'$, $t' \in [0, t]$, i.e., $R_{t}^{f}(X',U')=\left\{\phi_{t}^{f}(x,\mathbf{u}(t))\in X : x\in X', u(t') \in U', t' \in [0, t]\right\}$. 
\textcolor{black}{As will be clearer in Section~3, the reachable set $R_{t}^{f}(X',U')$ will be useful to design a policy that meets our control objective. 
However, the exact computation of $R_{t}^{f}(X',U')$ is not analytically tractable for the system \req{system_f}, and thus some over-approximation techniques must be required. In this paper, we employ the notion of \textit{mixed-monotone} \cite{coogan2015} aiming to compute an over-approximation of $R_{t}^{f}(X',U')$, since this approach is potentially less conservative than other over-approximation techniques, such as those based on a Lipschitz continuity; for this clarification, see \rrem{rem:conservative}.} 
The definition of mixed-monotone is given as follows:

\begin{mydef}\label{def_mixedmonotone}
\normalfont
Consider the system $dx(t)/dt = f(x(t), u(t))$ with the state $x \in X \subset \mathbb{R}^n$ and the control input $u \in U \subset \mathbb{R}^m$.
The system is \textit{mixed-monotone} if there exists a Lipschitz continuous function $d = [d_1, \ldots, d_n]^\top: X\times U \times X \times U\rightarrow X$, such that all of the following conditions hold:
\renewcommand{\labelenumi}{(\roman{enumi})}
\begin{enumerate}
    \item For all $x\in X$ and all $u\in U$, $d(x,u,x,u)=f(x,u)$.
    \item For all $i,j\in \left\{1,...,n\right\}$ with $i\ne j$, $\frac{\partial d_i}{\partial x_j}(x,u,\hat{x},\hat{u})\ge{0}$ for all $x,\hat{x}\in X$ and all $u,\hat{u}\in U$, where $x_j$ is the $j$-th element of $x$, whenever the derivative exists.
    \item For all $i,j\in \left\{1,...,n\right\}$ with $i\ne j$, $\frac{\partial d_i}{\partial \hat{x}_j}(x,u,\hat{x},\hat{u})\le{0}$ for all $x,\hat{x}\in X$ and all $u,\hat{u}\in U$, where $\hat{x}_j$ is the $j$-th element of $x$, whenever the derivative exists.
    \item For all $i\in \left\{1,...,n\right\}$ and $j \in \left\{1,...,m\right\}$, $\frac{\partial d_i}{\partial u_j}(x,u,\hat{x},\hat{u})\ge{0}$ and $\frac{\partial d_i}{\partial \hat{u}_j}(x,u,\hat{x},\hat{u})\le{0}$ for all $x,\hat{x}\in X$ and all $u,\hat{u}\in U$, where $u_j$ is the $j$-th element of $u$, whenever the derivative exists. \qedwhite 
\end{enumerate}
\end{mydef}

In general, the function $d$ in \rdef{def_mixedmonotone} is said to be a \textit{decomposition function}. Then, consider the following extended dynamical system: 
\begin{align}\label{extended_dynamics}
  \dot{\chi}(t)=  \left[
\begin{array}{c}
\dot{x}(t) \\
\dot{\hat{x}}(t)
\end{array}
\right] = \left[
\begin{array}{c}
d(x(t),u(t), \hat{x}(t), \hat{u}(t)) \\
d(\hat{x}(t), \hat{u}(t), {x}(t), {u}(t))
\end{array}
\right]. 
\end{align}
Given $\chi = [{x},\hat{x}]^\top \in X\times X$ and $v = [{u},\hat{u}]^\top \in U\times U$, let 
\begin{align}\label{chi_t}
\chi (t) = [\phi_{t}^{d}(({x},\hat{x}),({u},\hat{u})), \hat{\phi}_{t}^{d}(({x},\hat{x}),({u},\hat{u}))]^\top \in X \times X
\end{align}
be the solution of \req{extended_dynamics} starting from the state $\chi = [{x},\hat{x}]^\top$ with the control input $v = [{u},\hat{u}]^\top$ applied for $t\ge{0}$. That is, $\phi_{t}^{d}(({x},\hat{x}),({u},\hat{u}))$ and $\hat{\phi}_{t}^{d}(({x},\hat{x}),({u},\hat{u}))$ are the solution of $x(t)$ and $\hat{x}(t)$ by applying the constant control input $v = [{u},\hat{u}]^\top$ in \req{extended_dynamics}, respectively. For brevity, we denote $\phi_{t}^{d}(({x},\hat{x}),$ $({u},{u})) = \phi_{t}^{d}(({x},\hat{x}),u)$. In addition, let ${\phi}_{t,S}^{d}(\cdot)$ and ${\phi}_{t,I}^{d}(\cdot)$ (resp. $\hat{\phi}_{t,S}^{d}(\cdot)$ and $\hat{\phi}_{t,I}^{d}(\cdot)$) denote the first and the second element of ${\phi}_{t}^{d}(\cdot)$ (resp. $\hat{\phi}_{t}^{d}(\cdot)$), respectively. 

The mixed monotonicity property is useful since it gives an over-approximation of the reachable set as follows: 

\begin{myprop}\label{reachableapproxprop}
\normalfont
Consider the system $dx(t)/dt = f(x(t), u(t))$ with the state $x \in X \subset \mathbb{R}^n$ and the control input $u \in U \subset \mathbb{R}^m$, and suppose that it is mixed-monotone with the decomposition function $d$. 
Then, for all $X' = \llbracket \underline{x}, \bar{x} \rrbracket \subseteq X$, $U' = \llbracket \underline{u}, \bar{u} \rrbracket \subseteq  U$ and $t \geq 0$, it follows that 
\begin{align}\label{reachable_set}
    R_{t}^{f}(X', U')\subseteq \llbracket \phi_{t}^{d}((\underline{x},\bar{x}),(\underline{u},\bar{u})),\  \hat{\phi}_{t}^{d}((\underline{x},\bar{x}),(\underline{u},\bar{u})). \rrbracket 
\end{align}
\qedwhite 
\end{myprop}

For the proof, see \cite{coogan2015}. 
Following \cite{coogan2015}, let us introduce a decomposition function for \req{system_f}. Let $d = [d_1, d_2]^\mathsf{T}$ be the function defined by 
\begin{align}
& d_{1}(x,u,\hat{x},\hat{u})=-\hat{u}\hat{S} \hat{I}+\textcolor{black}{\xi(1-\hat{S}-\hat{I})},\label{d1}\\
& d_{2}(x,u,\hat{x},\hat{u})=
\begin{cases}
(uS-\gamma)I, & \mbox{if }uS/\gamma\ge{1} \\
(uS-\gamma)\hat{I}, & \mbox{if }uS/\gamma<{1}, \label{d2}
\end{cases}
\end{align}
where $x = [S, I]^\top$ and $\hat{x} = [\hat{S}, \hat{I}]^\top$. 
The following result shows that the function $d$ defined above is indeed a decomposition function for \req{system_f}: 
\begin{myprop}
\normalfont
Let $d = [d_1, d_2]^\mathsf{T}$ be the function given by \req{d1}, \req{d2}. Then, the system \req{system_f} is mixed monotone with the decomposition function $d$.  \qedwhite 
\end{myprop}

\noindent
\textbf{Proof}: 
Condition~(i) in \rdef{def_mixedmonotone} holds since $d_1(x,u,x,u) = - u S I+\textcolor{black}{\xi(1-S-I)}$ and $d_2 (x,u,x,u) = uSI - \gamma I$, which correspond to \req{state_s} and \req{state_i}, respectively. To check condition~(ii), note that we have 
\begin{align} \label{dx1}
\frac{\partial d}{\partial x}=
\begin{bmatrix}
\frac{\partial d_1}{\partial S} & \frac{\partial d_2}{\partial S} \\ \\
\frac{\partial d_1}{\partial I} & \frac{\partial d_2}{\partial I} 
\end{bmatrix}=
\begin{cases}
\begin{bmatrix}
0 & uI \\
0 & uS-\gamma 
\end{bmatrix}, & \mbox{if }uS/\gamma\ge{1} \\ \\
\begin{bmatrix}
0 & u\hat{I} \\
0 & 0  
\end{bmatrix}, & \mbox{if }uS/\gamma<1. 
\end{cases}
\end{align}
 Since $S,I,u$ are all non-negative, all the components of the derivatives in \req{dx1} are non-negative, and hence condition~(ii) holds. Similarly, condition~(iii) holds since  
\begin{align} \label{dx2}
\frac{\partial d}{\partial \hat{x}}=
\begin{bmatrix}
\frac{\partial d_1}{\partial \hat{S}} & \frac{\partial d_2}{\partial \hat{S}} \\ \\
\frac{\partial d_1}{\partial \hat{I}} & \frac{\partial d_2}{\partial \hat{I}}
\end{bmatrix}=
\begin{cases}
\begin{bmatrix}
-\hat{u}\hat{I}\textcolor{black}{-\xi} & 0 \\
-\hat{u}\hat{S}\textcolor{black}{-\xi} & 0 
\end{bmatrix}, & \mbox{if }uS/\gamma\ge{1} \\ \\
\begin{bmatrix}
-\hat{u}\hat{I}\textcolor{black}{-\xi} & 0 \\
-\hat{u}\hat{S} & uS-\gamma 
\end{bmatrix}, & \mbox{if }uS/\gamma<1 
\end{cases}
\end{align}
 and all the components in \req{dx2} are non-positive. 
 Finally, the condition (iv) holds since $\frac{\partial d_1}{\partial u} = [0, SI]^\mathsf{T}$ or $[0, S\hat{I}]^\mathsf{T
 }$ and $\frac{\partial d_1}{\partial \hat{u}} = [-\hat{S}\hat{I}, 0]^\mathsf{T}$ or $[-{S}\hat{I}, 0]^\mathsf{T}$, and thus all the components of the derivatives with respect to $u$ and $\hat{u}$ are non-negative and non-positive, respectively. Therefore, it is shown that the system \req{system_f} is mixed monotone with the decomposition function $d$. The proof is complete. \qedwhite

\smallskip 
For simplicity, we assume in this paper that there exist no uncertainties on the control input, which means that the over-approximation of the reachable set will be obtained (and simplified) as 
\begin{align}\label{reachablesimple}
    R_{t}^{f}(X', u)\subseteq \llbracket \phi_{t}^{d}((\underline{x},\bar{x}),u),\  \hat{\phi}_{\textcolor{black}{t}}^{d}((\underline{x},\bar{x}),u) \rrbracket. 
\end{align}
Hence, \req{reachablesimple} is utilized to obtain our result, as will be seen in the next section. 
However, our problem setup and the proposed approach can easily be extended to the case where we have uncertainties on the control input based on \req{reachable_set}. 
\textcolor{black}{As will be discussed later in the next section, over-approximation of the reachable set according to \req{reachablesimple} is useful to construct the symbolic model, which serves as the abstraction of the original SIRS model; for details, see in particular \rdef{symbolic model} in \rsec{solution approach}.} 

\begin{figure*}[t]
    \centering
    \includegraphics[width=0.8\textwidth]{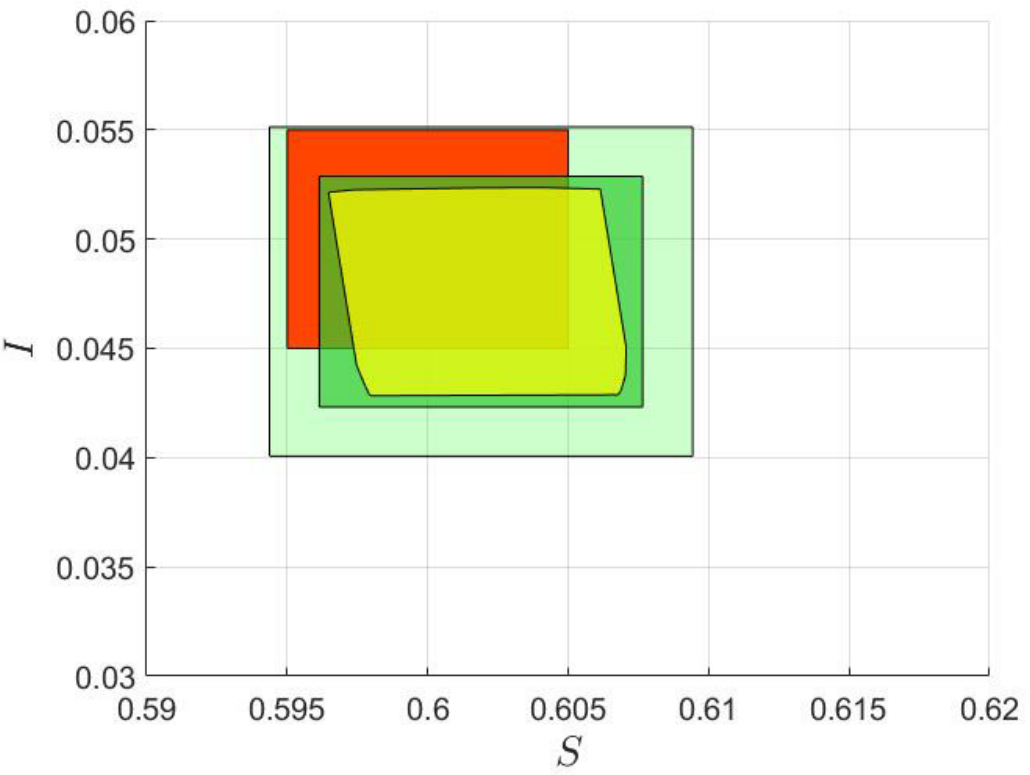}
    \caption{\textcolor{black}{Example of the over-approximations for $R_{t}^{f}(X',U')$. Here we assume $X'=[0.595,0.605]\times [0.045,0.055]$  and $U'=\left\{0.17\right\}, t=1$. $X'$ is shown in red, the true reachable set $R_{t}^{f}(X',u)$ (with $t=1, u=0.17$) is shown in yellow, and the over-approximations obtained from the mixed monotonicity and the Lipschitz continuity are shown in green and light green, respectively.}}
    \label{lipschitz}
\end{figure*}

\begin{remark}\label{rem:conservative}
\normalfont 
 \textcolor{black}{An alternative way to compute the over-approximation of the reachable set $R_{t}^{f}(X',U')$ is to use the Lipschitz continuity property of \req{system_f}. 
 Since $X$ is compact and closed, there exist $L_f >0$ such that $\|f(x_1, u) - f(x_2, u)\|_{\infty} \leq L_f \|x_1 - x_2\|_{\infty} $ for all $x \in X$ and $u \in U$. Hence, by using the Gronwall-Bellman inequality, we have $R^f _t (X', u) \subseteq \mathcal{B} (\phi^f _t (x', u), \varepsilon e^{L_f t})$, with $X' = \mathcal{B}(x', \varepsilon)$ {(i.e., the initial state starts from the square with the center $x'$ and the side length 2$\varepsilon$).} However, since the Lipschitz constant $L_f$ is the worst case upper bound of the derivative of $f$ (w.r.t $x$), the corresponding over-approximation of the reachable set may become more conservative than the one with the mixed monotonicity. To illustrate this, \rfig{lipschitz} shows a simple simulation result of $R_{t}^{f}(X',U')$ computed by the mixed monotonicity (green) and by the Lipschitz continuity (light green) for the SIRS model \req{system_f}. The figure shows that the (over-approximated) region by the Lipschitz continuity method is larger than the one by the mixed monotonicity method. \qedwhite}
\end{remark}


\subsection{Event-triggered control}\label{event-triggered control}
In order to deal with the nature of updating the control input according to the increase/decrease of the infected subjects, we formalize the \textit{event-triggered} control framework. 
In the event-triggered control framework, the time instants of updating the control inputs are determined by an event-triggering condition, which is defined by setting a certain threshold with respect to the change of the fraction of infected subjects $I$, i.e., only if $I$ changes over the threshold, the control input is updated. 
\textcolor{black}{The \textit{event-triggered control policy} in this paper is denoted as $\Pi = \{\pi_0, \pi \}$, where
\begin{align}
  \pi_0 : X_0 \rightarrow 2^{U\times E}, \   \pi : X \rightarrow 2^{U \times E}, 
\end{align}
with $E = [0,1)$ being the range of the thresholds.}
The {triggering time instants} (i.e., the time instants when updating control inputs), as well as the control inputs, are determined according to the policies $\pi_0, \pi$ as follows. Let the triggering time instants be given by $t_k$, $k=0, 1, 2, \ldots$ with $t_0 = 0$ and $t_k < t_{k+1}$ for all $k=0, 1, \ldots$. Then, 
\begin{align} 
&t_{k+1}=\mathrm{inf} \left\{t>t_k:|I(t)-I(t_k)| \geq \varepsilon (t_k) \right\}, \label{update} \\ 
&u(t) = u(t_k),\ t \in [t_k, t_{k+1}),\label{u}
\end{align}
 where $\{u(t_0), \varepsilon(t_0)\} \in \pi_0 (x(t_0))$ and $\{u(t_k), \varepsilon(t_k)\} \in \pi(x(t_k))$ for all $k=1, 2, \ldots$. 
 That is, for each $k=0, 1, \ldots $, the control input $u(t_k)$ is applied constantly until the next triggering time instant $t_{k+1}$ when $I(t)$ changes over the threshold $\varepsilon(t_k)$, and both $u(t_k)$ and $\varepsilon(t_k)$ are determined according to $\pi_0$ for the initial time instant ($k=0$) and $\pi$ otherwise ($k >0$).  
 \textcolor{black}{The main reason for differentiating two control policies $\pi_0, \pi$ is that it is necessary to define different transition functions for characterizing the abstracted behavior of the dynamical system according to the initial time and the non-initial time (for details, see \rsec{symbolic approach}).} 
Given $\Pi = \{\pi_0, \pi \}$ and $x \in X$, we denote by $\Phi^f_{\Pi,t}(x) \subset X$, $t \geq 0$ {the set of all states of the system} \req{system_f} which are possible to be reached at $t$ by applying the event-triggered control policies $\Pi = \{\pi_0, \pi \}$ starting from $x(t_0) = x$. 

\subsection{Control objective}\label{control objective}
\textcolor{black}{Our goal is to devise control strategies that meet the following critical objectives: 
(i)~\textit{ICU Availability:} It is essential to consistently keep the fraction of infected subjects under a specific critical level, denoted by $\bar{I}_S$, to maintain adequate Intensive Care Units (ICUs), i.e., $I(t) \leq \bar{I}_S$ for all $t \geq 0$;
{\color{black}(ii)~\textit{Vulnerable Population Protection:} Given the high severity rate among groups with weaker immune systems like the elderly and children post-infection, we aim to ensure a substantial fraction of susceptible subjects, i.e., $S(t) \geq \underline{S}_S$ for a given $\underline{S}_S$ for all $t \geq 0$; 
(iii)~\textit{Pandemic Containment:} 
We strive to ensure that the fraction of infected subjects falls below a specific threshold and the fraction of susceptible subjects is greater than a specific level $\underline{S}_F$ within a finite time period, i.e., for every $x(t_0) \in [S(t_0), I(t_0)]^\top \in X_0$, there exists $t'$ such that $I(t') \leq \bar{I}_F$ and $S(t') \geq \underline{S}_F$ for given $\bar{I}_F$, $\underline{S}_F$;
(iv)~\textit{Reinfection Prevention} : To avert a secondary outbreak caused by reinfection, once the condition in (iii) is achieved at $t'$, we aim to keep it satisfied for all future times, i.e., we aim to ensure that $I(t)\leq \bar{I}_F,S(t)\geq \underline{S}_F$, for all $t\geq t'$; (v) finally, in order not to restrict the freedom of economic activity, the total cost to execute the intervention (i.e., control input) is to be minimized.}}

{\color{black}To specifically formalize the control objectives for (i)--(iv), let $X_S, X_F \subset X$
be given by 
\begin{align}\label{xsdef}
    X_S &= \{x = [S, I]^\mathsf{T}\in X: S \geq \underline{S}_S\ \mathrm{and}\ I \leq \bar{I}_S\}, \\
    X_F & = \{x = [S, I]^\mathsf{T}\in X: S \geq \underline{S}_F\ \mathrm{and}\ I \leq \bar{I}_F\},\label{xfdef}
\end{align}
for given $\underline{S}_S, \underline{S}_F, \bar{I}_S, \bar{I}_F \in (0, 1)$ with $\underline{S}_F>\underline{S}_S,\bar{I}_F < \bar{I}_S$.} We assume that $X_0 \subseteq X_S$ and $X_F \cap X_0 = \varnothing$. From the above, $X_F \subset X_S$ trivially holds. Then, we define the notion of a \textit{valid} pair of the control policies as follows: 
\begin{mydef}[Valid \textit{reachable} event-triggered control policy]\label{terminal control policy}
\normalfont
Given $U$, $X_0$, $X_S$ and $X_F$, we say that $\Pi =\{\pi_{0}, \pi\}$ is a \textit{valid} pair of the \textit{reachable} event-triggered {control policy} if the following holds: 
for every $x(t_0)\in X_0$, there exists \textcolor{black}{$t'\in [t_0,\infty)$} such that $\Phi^f_{\Pi,t'}(x(t_0))\subseteq X_F$ and $\Phi^f_{\Pi,t}(x(t_0))\subseteq X_S$ for all $t\in [t_0, t']$. 
\qedwhite 
\end{mydef}
That is, $\Pi$ is a valid {reachable} event-triggered control policy if for any $x(t_0)\in X_0$, {any} controlled trajectory starting from $x(t_0)$ by applying $\Pi$ reaches $X_F$ in finite time while remaining in $X_S$. 
\begin{mydef}[Valid \textit{terminal} event-triggered control policy]\label{safe control policy}
\normalfont
Given $U$, $X_S$ and $X_F$, we say that $\pi_F$ is a valid \textit{terminal} event-triggered {control policy}
if the following holds: $x(t_0)\in X_F \implies \Phi^f_{\Pi_F,t}(x(t_0))\subseteq X_F$ for all $t\in [t_0, \infty]$. 
\qedwhite 
\end{mydef}
That is, $\pi_F$ is a valid terminal event-triggered control policy if \textit{any} controlled trajectory starting from $X_F$ remains in $X_F$ for all times. 
Hence, by designing valid reachable and terminal event-triggered control policies, we can ensure that any controlled trajectory starting from any $x(t_0) \in X_0$ reaches $X_F$ in finite time (by applying $\Pi$), and then stays therein for all future times {(by applying $\pi_F$)}. 
The problem that we seek to solve for the objectives (i)--(iv) given above is to synthesize a valid pair of the above event-triggered control policies. 
 \begin{mypro}\label{problem1}
 \normalfont 
 Given $U$, $X_0$, $X_F$ and $X_S$, synthesize both the reachable and the terminal event-triggered control policies.  \qedwhite 
 \end{mypro}

 The control policy for additionally achieving the objective (v) will be elaborated in more details after providing the solution to \rpro{problem1} (\rsec{selectionsec}). 

\section{Solution Approach}\label{solution approach}
In this section, we provide a solution approach to \rpro{problem1}.

\subsection{Event-triggered transition system}
\label{event-triggered_transition_system}
Let us first describe the SIRS model \req{system_f} by a \textit{transition system} as follows: 
\begin{mydef}\label{transitionSIR}
\normalfont 
A transition system for the SIRS model \req{system_f} is a tuple $T=(X,X_0,U,f)$, where:
\begin{itemize}
    \item $X$ is a set of states;
    \item $X_0\subset X$ is a set of initial states; 
    \item $U= \{u_0, \ldots, u_{m-1}\}$ is a set of control inputs;
    \item $f:X\times U \rightarrow X$ is a function describing the dynamics of the SIRS model \req{system_f}. \qedwhite
\end{itemize}
\end{mydef}

\textcolor{black}{Now, we will adapt 
\rdef{transitionSIR} to include the event-triggered strategy within the notion of the transition system.} 
As stated in the previous section, the control input is applied constantly until the next updating time instant when the fraction of infected subjects changes over a certain threshold (see \req{update}). 
In this paper, we express this fact by introducing the \textit{event-triggered transition system} of $T$, which is formally defined as follows:
\begin{mydef}\label{transitionEV}
\normalfont 
Let $T$ be a transition system for the SIRS model \req{system_f}. 
The \textit{event-triggered transition system} of $T$ is a tuple $T_{e}=(X,X_0,U,E,g)$, where:
\begin{itemize}
    \item $X$ is a set of states;
    \item $X_0 \subset X$ is a set of initial states;
\item $U = \{u_0, \ldots, u_{m-1}\}$ is a set of control inputs;
\item $E = [0,1)$ is the range of the thresholds;
\item $g: X\times U\times E\rightarrow {X}$ is a {transition function}, where $x' = g(x,u,\varepsilon)$ with $x = [S, I]^\mathsf{T} \in {X}$, $x' = [S^{'}, I^{'}]^\mathsf{T} \in {X}$, $u \in U, \varepsilon \in E$ iff the following condition holds: 
$I^{'} >0$, $\varepsilon >0$ and $x'= \phi_{t^{'}}^{f}(x,u)$, where 
    \begin{align}\label{delta}
    &t' =\mathrm{inf}\left\{t >0:|\phi_{t,I}^{f}(x,u)-I|\ge{\varepsilon}\right\}
    \end{align}
    \qedwhite
\end{itemize}
\end{mydef}

The notion of the event-triggered transition system (\rdef{transitionEV}) differs from the original one (\rdef{transitionSIR}), in the sense that we introduce the transition function $g$ that captures the transition of the states according to the event-triggered strategy as formalized in \rsec{event-triggered control}. 
Specifically, if we have the relation $x' = g(x, u, \varepsilon)$ satisfying the condition in 
\rdef{transitionEV}, 
it means that the state is steered from $x$ to $x'$ by applying the constant control input $u$ until the fraction of infected subjects changes by the threshold $\varepsilon$, 
i.e., letting $x=[S, I]^\top$ and $x' = [S^{'},I^{'}]^\top$, it follows either $I^{'} = I+\varepsilon$ or $I^{'} = I-\varepsilon$. 

\subsection{Symbolic model and approximate alternating simulation relation}\label{symbolic approach}
Based on the event-triggered transition system obtained in the previous section, let us now provide a \textit{symbolic model}, which captures abstracted behaviors of $T_e$ and allows us to synthesize a valid pair of the event-triggered control policies. 

\begin{mydef}\label{symbolic model}
\normalfont 
Let $T_{e}$ be the event-triggered transition system in \rdef{transitionEV}. 
Given $\eta_{S}, \eta_I>0$, a \textit{symbolic model} of $T_{e}$ is a tuple $\Tilde{T}_{\eta_S,\eta_I} =(\Tilde{X},\Tilde{X}_0,\tilde{U},\tilde{E},\Tilde{g}, \tilde{g}_0, L)$, where:
\begin{itemize}
    \item $\Tilde{X}=[X]_{\eta_{S}, \eta_I}$ is a set of states;
    \item $\Tilde{X}_0=[\mathrm{Out}_{\eta_S/2, \eta_I/2} (X_0)]_{\eta_{S}, \eta_I}$ is a set of initial states;
\item $\tilde{U} = U = \{u_0, \ldots, u_{m-1}\}$ is a set of control inputs;
\item $\tilde{E}= \left\{\varepsilon_0,...,\varepsilon_{n-1}\right\} \cup \{0\}$ with $0< \varepsilon_0<...<\varepsilon_{n-1} <1$ is a set of the candidate thresholds, such that for every $i \in \{0, \ldots, n-1\}$, there exists  $k_i \in \mathbb{Z}$ satisfying $\varepsilon_i = k_i \eta_I$;
\item $\Tilde{g}:\Tilde{X}\times \tilde{U}\times \tilde{E}\rightarrow2^{\Tilde{X}}$ is a transition function for the \textit{non-initial time}, which is defined as follows. 
For given $\Tilde{x} = [\Tilde{S}, \Tilde{I}]^\mathsf{T} \in \tilde{X}$, $\Tilde{x}' = [\Tilde{S}^{'}, \Tilde{I}^{'}]^\mathsf{T} \in \tilde{X}$, $\Tilde{u} \in \tilde{U}, \Tilde{\varepsilon} \in \tilde{E}$, let $\underline{\tilde{x}} = [\tilde{S}-\eta_S/2, \tilde{I}]^\mathsf{T}$, $\bar{\tilde{x}} = [\tilde{S}+\eta_S/2, \tilde{I}]^\mathsf{T}$ and 
\begin{align}
        \bar{t} &= \inf \{t >0: {\phi}_{{t},\tilde{I}}^{d}((\underline{\tilde{x}},\bar{\tilde{x}}), \Tilde{u}) = \Tilde{I}^{'}\}, \label{overline-t}\\ 
        \underline{t} &= \inf \{t >0: \hat{\phi}_{{t},\tilde{I}}^{d}((\underline{\tilde{x}},\bar{\tilde{x}}), \Tilde{u}) = \Tilde{I}^{'}\}. \label{underline-t}
    \end{align}
Then, $\Tilde{x}' \in \Tilde{g}(\Tilde{x},\Tilde{u},\Tilde{\varepsilon})$ iff {one} of the following conditions holds: 
\renewcommand{\labelenumi}{(\roman{enumi})}
\begin{enumerate}
    \item $\tilde{I}>0$, $\tilde{\varepsilon}>0$, $\Tilde{I}^{'}=\tilde{I}+\tilde{\varepsilon} >0$, $ \bar{t} < \infty$, $\underline{t} < \infty$, \textcolor{black}{${\phi}_{{t},S}^{d}((\underline{\tilde{x}},\bar{\tilde{x}}), \tilde{u}) \geq \underline{S}+\eta_S/2$, $\forall t \in [0, \underline{t}]$,} and \textcolor{black}{there exists $t \in [\bar{t}, \underline{t}]$ such that  
    \begin{align}
       \tilde{S}^{'}\in \left[\hat{\phi}_{t,S}^{d}((\underline{\tilde{x}},\bar{\tilde{x}}), \tilde{u})-\frac{\eta_S}{2},{\phi}_{{t},S}^{d}((\underline{\tilde{x}},\bar{\tilde{x}}), \tilde{u})+\frac{\eta_S}{2}\right], \label{condition i}
    \end{align}}
    
    \item $\tilde{I}>0$, $\tilde{\varepsilon}>0$, $\Tilde{I}^{'}=\tilde{I}-\tilde{\varepsilon} > {0}$, $ \bar{t} < \infty$, $\underline{t} < \infty$, \textcolor{black}{${\phi}_{{t},S}^{d}((\underline{\tilde{x}},\bar{\tilde{x}}), \tilde{u}) \geq \underline{S}+\eta_S/2$, $\hat{\phi}_{{t},I}^{d}((\underline{\tilde{x}},\bar{\tilde{x}}), \tilde{u}) < \tilde{I}+\tilde{\varepsilon}$, $\forall t \in [0, \bar{t}]$} and \textcolor{black}{there exists $t \in [\underline{t}, \bar{t}]$ such that 
    \begin{align}
         \tilde{S}^{'}\in \left[\hat{\phi}_{{t},S}^{d}((\underline{\tilde{x}},\bar{\tilde{x}}), \tilde{u})-\frac{\eta_S}{2},{\phi}_{{t},S}^{d}((\underline{\tilde{x}},\bar{\tilde{x}}), \tilde{u})+\frac{\eta_S}{2}\right], \label{condition ii}
    \end{align}}
\end{enumerate}
\item $\Tilde{g}_0:\Tilde{X}_0 \times \tilde{U}\times \tilde{E}\rightarrow2^{\Tilde{X}}$ is a transition function for the initial time, which is defined in the same way as $\Tilde{g}$, except that we have $\underline{\tilde{x}} = [\tilde{S}-\eta_S/2, \tilde{I}-\eta_I/2]^\mathsf{T}$, $\bar{\tilde{x}} = [\tilde{S}+\eta_S/2, \tilde{I}+\eta_I/2]^\mathsf{T}$. 
\item  $L : \Tilde{X}_0 \times \tilde{U}\times \tilde{E}\rightarrow \{0, 1\}$ is a labeling function, where $L(\Tilde{x},\Tilde{u},\Tilde{\varepsilon}) = 1$ iff $\tilde{I}' = \tilde{I} + \Tilde{\varepsilon}$ for all $\Tilde{x}' = [\tilde{S}', \tilde{I}']^\top \in \Tilde{g}_0 (\Tilde{x},\Tilde{u},\Tilde{\varepsilon})$ and $L(\Tilde{x},\Tilde{u},\Tilde{\varepsilon}) = 0$ iff $\tilde{I}' = \tilde{I} - \Tilde{\varepsilon}$ for all $\Tilde{x}' = [\tilde{S}', \tilde{I}']^\top \in \Tilde{g}_0 (\Tilde{x},\Tilde{u},\Tilde{\varepsilon})$.
\item \textcolor{black}{$L_F : \Tilde{X} \times \tilde{U}\times \tilde{E}\rightarrow \{0, 1\}$ is a labeling function for the terminal policy, where $L_F(\Tilde{x},\Tilde{u},\Tilde{\varepsilon}) = 1$ iff $\Tilde{g}(\Tilde{x},\Tilde{u},\Tilde{\varepsilon}) \neq \varnothing$, $\tilde{I}+\tilde{\varepsilon}\leq \bar{I}_F$ (with $\Tilde{x} = [\Tilde{S}, \Tilde{I}]^\mathsf{T}$) and
for every $\Tilde{x}' \in \Tilde{g}(\Tilde{x},\Tilde{u},\Tilde{\varepsilon})$ we have
${\phi}_{{t},S}^{d}((\underline{\tilde{x}},\bar{\tilde{x}}), \tilde{u}) \geq \underline{S}_F+\eta_S/2$, $\forall t \in [0, {\rm max}(\underline{t},\bar{t})]$. Otherwise, $L_F(\Tilde{x},\Tilde{u},\Tilde{\varepsilon}) = 0$.}
\qedwhite 
\end{itemize}
\end{mydef}
In contrast to \rdef{transitionEV}, the symbolic model deals with the \textit{discretized} sets $\tilde{X}$, $\tilde{X}_0$, $\tilde{U}$, $\tilde{E}$ and introduces the new transition functions $\tilde{g}$ and $\tilde{g}_0$ defined according to the conditions of (i),(ii). 
Note that each element in $\tilde{E}$ is chosen to be the integer multiple of $\eta_I$, so that it allows us to trigger the event always at some point on the lines $ I = j \eta_I$, $j \in \mathbb{Z}_{\geq 0}$. 
\textcolor{black}{We have here utilized different transition functions $\tilde{g}$, $\tilde{g}_0$, where $\tilde{g}$ captures the transitions for the non-initial time, and $\tilde{g}_0$ captures the transitions for the initial time, and these are constructed by computing the over-approximation of the reachable set according to \req{reachablesimple}.} 
Roughly speaking, $\tilde{g}_0$ is defined by computing an over-approximation of the reachable set of the transition system $T_e$ from a \textit{rectangle} (a subset of $X_0$) to a \textit{segment} (a set of states lying in  $I = j \eta_I$ for some $j \in \mathbb{Z}_{\geq 0}$). On the other hand, $\tilde{g}$ is defined by computing an over-approximation of the reachable set of $T_e$ from a \textit{segment} to a \textit{segment}. 
As mentioned above, this is because the event is triggered always on the lines for the non-initial time, i.e., $I (t_k) = j \eta_I$ for some $j \in \mathbb{Z}_{\geq 0}$ for all $k = 1, 2, ...$, while the initial state starts from any $x (t_0) = [S(t_0), I(t_0) ] ^\top \in X_0$ (i.e., the state at the initial time is not necessarily on the lines). 

\begin{figure}[t]
    \centering
    \includegraphics[width=1\textwidth]{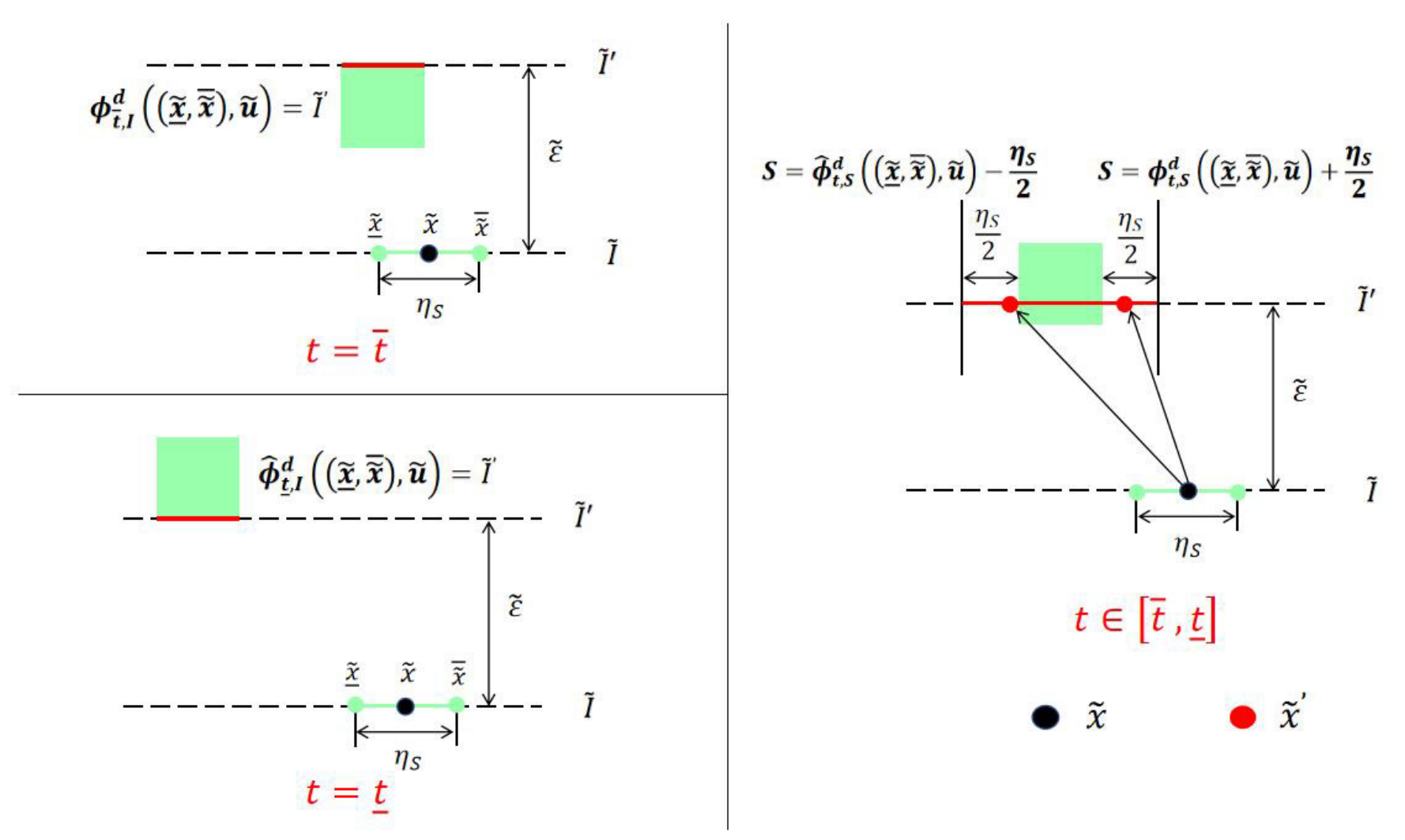}
    \caption{{Illustration of  condition~(i) in \rdef{symbolic model}. 
    The upper-left (resp. lower-left) figure illustrates the time instant $\bar{t}$ (resp. $\underline{t}$) that the upper-right point (resp. lower-left point) of the over-approximated reachable set (rectangle space in light green) crosses $\tilde{I}^{'}$. The right figure illustrates the transition function $\tilde{g}$, where the red points denote a set of successors from the discretized state represented by the black point according to the definition of $\tilde{g}$.}}
    \label{condition (i)}
\end{figure}

\begin{figure}[ht]
    \centering
    \includegraphics[width=1\textwidth]{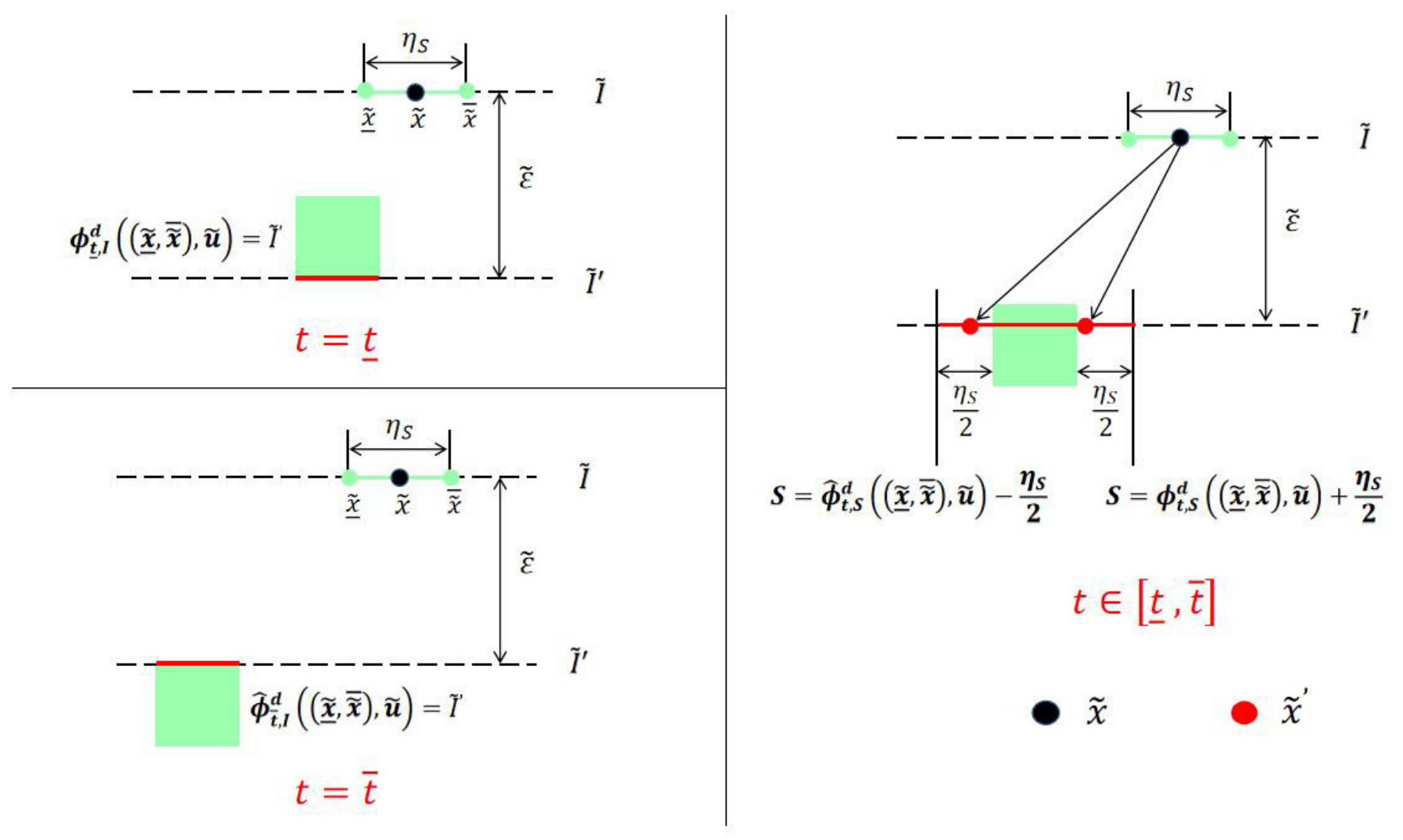}
    \caption{{Illustration of condition~(ii) in \rdef{symbolic model}. The upper-left (resp. lower-left) figure illustrates the time instant $\underline{t}$ (resp. $\overline{t}$) that the lower-left (resp. upper-right) point of the over-approximated reachable set (rectangle space in light green) crosses $\tilde{I}^{'}$. The right figure illustrates the transition function $\Tilde{g}$, where the red points denote a set of successors from the discretized state represented by the black point according to the definition of $\Tilde{g}$.}}
    \label{condition (ii)}
\end{figure}

More detailed explanations for defining $\tilde{g}$, $\tilde{g}_0$ are elaborated below. 
As in \req{overline-t}, \req{underline-t}, we first seek a time instant $\bar{t}$ when the upper-right point of the over-approximated reachable set, which is defined as ${\phi}_{{t},S}^{d}((\underline{\tilde{x}},\bar{\tilde{x}}),\tilde{u})$, crosses $\Tilde{I}^{'}$, and a time instant $\underline{t}$ when the lower-left point of the over-approximated reachable set, which is here defined as $\hat{\phi}_{{t},S}^{d}((\underline{\tilde{x}},\bar{\tilde{x}}), \tilde{u})$, crosses $\Tilde{I}^{'}$. 
In condition~(i), we find these instants for $\Tilde{I}^{'} = \Tilde{I} + \tilde{\varepsilon}$ 
(see \rfig{condition (i)} (left) for the illustration). 
The existence of finite $\bar{t} < \infty$ and $\underline{t} < \infty$ imply that every trajectory of \req{system_f} starting from $\llbracket \underline{\tilde{x}}, \bar{\tilde{x}} \rrbracket$ by applying $\tilde{u}$ must cross $\Tilde{I}^{'} = \Tilde{I} + \tilde{\varepsilon}$ for some $t \in [\underline{t}, \bar{t}]$, i.e., for every $x \in \llbracket \underline{\tilde{x}}, \bar{\tilde{x}} \rrbracket$, there exists $t \in [\underline{t}, \bar{t}]$ such that $\phi^f _{t,I} (x, u) = \tilde{I}+\tilde{\varepsilon}$.  
Hence, we add the transitions from $\tilde{x}=[\tilde{S}, \tilde{I}]^\top$ to $\tilde{x}^{'}=[\Tilde{S}^{'}, \Tilde{I}^{'}]^\top$ with $\Tilde{I}^{'} = \Tilde{I} + \tilde{\varepsilon}$ and all $\Tilde{S}^{'}$ contained in  the interval $[\hat{\phi}_{{t},S}^{d}((\underline{\tilde{x}},{\tilde{x}}),\tilde{u}) - \eta_S/2,{\phi}_{{t},S}^{d}((\underline{\tilde{x}},{\tilde{x}}), \tilde{u}) + \eta_S/2]$ for some $t \in [\underline{t}, \bar{t}]$. Hence, condition~(i) deals with the case when the event-triggering condition is violated when $I$ increases by $\tilde{\varepsilon}$ (i.e., $\Tilde{I}^{'} = \Tilde{I} + \tilde{\varepsilon}$). 
Similarly, condition~(ii) deals with the case when the event-triggering condition is violated at $\Tilde{I}^{'} = \Tilde{I} - \tilde{\varepsilon}$, i.e., we add the transitions from $\tilde{x}=[\tilde{S}, \tilde{I}]$ to $\tilde{x}^{'}=[\Tilde{S}^{'}, \Tilde{I}^{'}]$ with $\Tilde{I}^{'} = \Tilde{I} - \tilde{\varepsilon}$ and all $\Tilde{S}^{'}$ contained in the interval $[\hat{\phi}_{{t},S}^{d}((\underline{\tilde{x}},{\tilde{x}}), \tilde{u}) - \eta_S/2,{\phi}_{{t},S}^{d}((\underline{\tilde{x}},\bar{\tilde{x}}), \tilde{u}) + \eta_S/2]$ for some $t \in [\bar{t}, \underline{t}]$ (see \rfig{condition (ii)} for the illustration). 
The labeling function $L, L_F$ are utilized to synthesize the control policies (for details, see \rsec{control policy synthesis}). 

The symbolic model $\Tilde{T}_{\eta_S,\eta_I}$ constructed above represents an \textit{abstracted} expression of $T_e$, in the sense that it consists of only a finite number of states $\Tilde{X}=[X]_{\eta_{S}, \eta_I}$ (in contrast, $T_e$ consists of an infinite number of states $X$), while approximately mimicking the behaviors of $T_e$. 
More formally, the {similarities} between the two transition systems can be captured by the notion of an approximate alternating simulation relation defined as follows: 
\begin{mydef}\label{asrdef}
\normalfont 
Let $T_{e}=(X,X_0,U,E,g)$ and $\tilde{T}_{\eta_S,\eta_I}=(\Tilde{X},\Tilde{X}_0,\Tilde{U},\Tilde{E},\Tilde{g}, \Tilde{g}_0, L )$ be the event-triggered transition system and the corresponding symbolic model with the discretization parameters $\eta_S, \eta_I>0$, respectively. 
Then, a pair of the relations $(R_0, R)$ with $R_0 \subseteq \Tilde{X}_0 \times X_0$ and $R \subseteq \Tilde{X} \times X$ is called an $(\eta_{S}, \eta_I)$-approximate alternating simulation relation ($(\eta_{S}, \eta_I)$-ASR for short) from $\tilde{T}_{\eta_S,\eta_I}$ to $T_e$, if the following conditions hold: 
\renewcommand{\labelenumi}{(C\theenumi)}
\begin{enumerate}
    \item For every $\tilde{x}_{0} \in \tilde{X}_{0}$, there exists $x_{0}\in X_{0}$ such that $(\tilde{x}_{0}, x_{0})\in R_0$; 
    \item For every $(\tilde{x},x)\in R_0$, we have $| \Tilde{S}-S | \leq\eta_{S}/2$ and $| \Tilde{I}-I | \leq\eta_{I}/2$, where $x = [S, I]^\top, \tilde{x} = [\tilde{S}, \tilde{I}]^\top$. In addition, for every $(\tilde{x},x)\in R$, we have $| \Tilde{S}-S | \leq\eta_{S}/2$ and $\Tilde{I} = I$, where $x = [S, I]^\top, \tilde{x} = [\tilde{S}, \tilde{I}]^\top$; 
    \item For every $(\Tilde{x},x)\in R$ (resp. $(\Tilde{x},x)\in R_0$) and for every $(\tilde{u},\tilde{\varepsilon})\in \tilde{U}\times\tilde{E}$ with $\Tilde{g}(\Tilde{x},\Tilde{u},\Tilde{\varepsilon})\neq\varnothing$ (resp. $\Tilde{g}_0 (\Tilde{x},\Tilde{u},\Tilde{\varepsilon})\neq\varnothing$), there exists $({u}, {\varepsilon})\in U \times E$, such that: $x' = g(x,u,\varepsilon)$ implies the existence of $\tilde{x}' \in \Tilde{g}(\Tilde{x},\Tilde{u},\Tilde{\varepsilon})$ (resp. $\tilde{x}' \in \Tilde{g}_0(\Tilde{x},\Tilde{u},\Tilde{\varepsilon})$), such that $(\Tilde{x}',x')\in R$. \qedwhite  
\end{enumerate}
\end{mydef}

Note that the concept of an approximate alternating simulation relation defined above is slightly different from the standard one (e.g., \cite{tabuada2009verification}) in that two types of relations (i.e., $R_0$ and $R$) are defined, \textcolor{black}{which is because two different transition functions $\Tilde{g}_0$, $\tilde{g}$ are given in $\tilde{T}_{\eta_S,\eta_I}$. Specifically, $R_0$ is utilized to capture behavioral similarities for the \textit{initial time} (i.e., the transition functions between $g$ and $\Tilde{g}_0$), and $R$ is utilized to capture behavioral similarities for the \textit{non-initial time} (i.e., the transition functions between $g$ and $\Tilde{g}$).} 
The concept of $(\eta_{S}, \eta_I)$-ASR is useful for synthesizing policies as follows. Suppose that $\tilde{T}_{\eta_S,\eta_I}$ is constructed to guarantee the existence of a $(\eta_{S}, \eta_I)$-ASR from $\tilde{T}_{\eta_S,\eta_I}$ to $T_e$. Then, it is shown that the existence of a control policy for $\tilde{T}_{\eta_S,\eta_I}$ satisfying a given specification implies the existence of a control policy for $T_e$ satisfying the same specifications, i.e., the control policy that is synthesized for the symbolic model can be \textit{refined} to the control policy that can be applied to the original system \cite{reissig2016feedback}. 

The existence of a $(\eta_{S}, \eta_I)$-ASR from $\tilde{T}_{\eta_S,\eta_I}$ to $T_e$, can be justified by the following result:
\begin{mythm}\label{lemma1}
\normalfont
 Let $T_{e}=(X,X_0,U,E,g)$ and $\Tilde{T}_{\eta_S,\eta_I} =(\Tilde{X},\Tilde{X}_0,\Tilde{U},\Tilde{E},\Tilde{g}, \tilde{g}_0, L)$ be the event-triggered transition system and the corresponding symbolic model for given discretization parameters $\eta_S, \eta_I >0$, respectively. Let 
\begin{align}
&R_0=\left\{\left ([\tilde{S},\tilde{I}]^\top,[S, I]^\top \right) \in  \tilde{X}_0 \times X_0 : |\Tilde{S}-S| \leq\eta_{S}/2,\ |\Tilde{I}-I| \leq\eta_{I}/2 \right\}, \label{asr_r0} \\ 
&R=\left\{\left ([\tilde{S},\tilde{I}]^\top,[S, I]^\top \right) \in \tilde{X} \times X: |\Tilde{S}-S| \leq\eta_{S}/2,\ \Tilde{I} = I \right\}. \label{asr_r}
\end{align}

 Then, the pair of the relations $(R_0, R)$ is $(\eta_{S}, \eta_I)$-ASR from $\Tilde{T}_{\eta_S,\eta_I}$ to $T_{e}$. \qedwhite 
\end{mythm}
\noindent 
$\mathbf{Proof. }$ Let $R_0$ and $R$ be given by \req{asr_r0} and \req{asr_r}. 
Since $\Tilde{X}_{0} = [\mathrm{Out}_{\eta_S/2, \eta_I/2} $ $ (X_0)]_{\eta_{S}, \eta_I}$, for every $\Tilde{x}_0 =[\tilde{S}_0,\tilde{I}_0]^\top\in \Tilde{X}_{0}$, there exists $x_0=[S_0,I_0]^\top\in X_{0}$ such that $|\tilde{S}_0-S_0| \leq \eta_S/2$ and  $|\tilde{I}_0-I_0| \leq \eta_I/2$, i.e., $(\tilde{x}_0, x_0)\in R_0$. Hence, (C1) in \rdef{asrdef} holds. 
The condition of (C2) is directly satisfied from the definition of $R_0$ and $R$. 
Let us now check (C3). 
We first show that (C3) holds with the transition function $\tilde{g}$. 
Consider any $\tilde{x}=[\tilde{S},\tilde{I}]^\top\in X,(\tilde{u},\tilde{\varepsilon})\in \tilde{U}\times \tilde{E}$ with $\Tilde{g}(\Tilde{x},\Tilde{u},\Tilde{\varepsilon})\neq\varnothing$. For any $x=[S,I]^\top\in X$, $(\tilde{x},x)\in R$, let $u=\tilde{u}$, $\varepsilon=\tilde{\varepsilon}$, and $x'=g(x,u,\varepsilon)$ with $x'=[S^{'},I^{'}]^\top$. First, consider the transition for the case $I' = I+ \varepsilon$ (i.e., the fraction of infected subjects increases by $\varepsilon$). 
Pick $\tilde{x}'=[\tilde{S}^{'},\tilde{I}^{'}]^\top\in {\rm Nearest}_{\Tilde{X}}(x')$. 
We have $|\tilde{S}^{'}-S^{'}|\leq\eta_{S}/2$ and $\tilde{I}^{'}=I^{'}$ and thus $(\tilde{x}',x')\in R$. Since $x\in \llbracket\underline{\tilde{x}},\bar{\tilde{x}}\rrbracket$ with $\underline{\tilde{x}},\bar{\tilde{x}}$ being defined in \rdef{symbolic model} and using \rprop{reachableapproxprop} and the decomposition function $d$ as obtained in \req{d1} and \req{d2},  
there exists $t$ such that
$S^{'}\in [\hat{\phi}_{t,S}^{d}((\underline{\tilde{x}},\bar{\tilde{x}}), \tilde{u}),{\phi}_{t,S}^{d}((\underline{\tilde{x}},\bar{\tilde{x}}), \tilde{u})]$. Hence, $\tilde{S}^{'}\in [\hat{\phi}_{{t},S}^{d}((\underline{\tilde{x}},\bar{\tilde{x}}), \tilde{u})-\eta_S/2,{\phi}_{{t},S}^{d}((\underline{\tilde{x}},\bar{\tilde{x}}), \tilde{u})+\eta_S/2]$ and thus $\tilde{x}'\in \tilde{g}(\tilde{x},\tilde{u},\tilde{\varepsilon})$ according to \req{condition i} (i.e., condition~(i) in \rdef{symbolic model}), so that (C3) in \rdef{asrdef} holds. 
The transitions for the case $I' = I -\varepsilon$ (the fraction of infected subjects decreases by $\varepsilon$), which corresponds to condition~(ii) in \rdef{symbolic model}, can be proven in the same way as above and is thus omitted for brevity. 
Hence, it is shown that (C3) holds for the transition function $\tilde{g}$. 

Let us now show that (C3) holds with the transition function $\tilde{g}_0$. Consider any {$\tilde{x}=[\tilde{S},\tilde{I}]^\top\in \tilde{X}_0,(\tilde{u},\tilde{\varepsilon})\in \tilde{U}\times \tilde{E}$} with $\Tilde{g}_0 (\Tilde{x},\Tilde{u},\Tilde{\varepsilon})\neq\varnothing$. For any {$x=[S,I]^\top\in X_0$, $(\tilde{x},x)\in R_0$}, let $u=\tilde{u}$ and $\varepsilon=\tilde{\varepsilon}+ (\tilde{I} - I)$ (if $L(\Tilde{x},\Tilde{u},\Tilde{\varepsilon}) = 1$) or $\varepsilon=\tilde{\varepsilon} + (I-\tilde{I})$ ($L(\Tilde{x},\Tilde{u},\Tilde{\varepsilon}) = 0$), and $x'=g(x,u,\varepsilon)$ with $x'=[S^{'},I^{'}]^\top$. Then, by following exactly the same procedure as for the case $\tilde{g}$, it is shown that there exists $\tilde{x}' \in \tilde{g}_0 (\tilde{x}, \tilde{u}, \tilde{\varepsilon})$ such that $(\tilde{x}', x') \in R$ holds. 
Therefore, $(R_0, R)$ is $(\eta_S, \eta_I)$-ASR from $\Tilde{T}_{\eta_S,\eta_I}$ to $T_e$. \qedwhite

\subsection{\textcolor{black}{Control policy synthesis}}\label{control policy synthesis}
In this section, we present an algorithm to synthesize a valid pair of the terminal and the reachable event-triggered control policies as the solution to \rpro{problem1}. Given $T_e$, suppose that $\Tilde{T}_{\eta_S,\eta_I}$ is constructed such that the pair $(R_0, R)$ is $(\eta_S, \eta_I)$-ASR from $\Tilde{T}_{\eta_S,\eta_I}$ to $T_e$, where $(R_0, R)$ is defined in \rthm{lemma1}. We first synthesize the policies for $\Tilde{T}_{\eta_S,\eta_I}$. Then, we synthesize the policies for $T_e$. 


Let $X_S, X_F$ be the sets as defined in \rsec{control objective} and $\Tilde{X}_S=[{\rm lnt}_{\eta_S,\eta_I}(X_S)]_{\eta_S,\eta_I}$, $\Tilde{X}_F=[{\rm lnt}_{\eta_S,\eta_I}(X_F)]_{\eta_S,\eta_I}$. This implies that
\begin{align}
&\Tilde{x}\in \Tilde{X}_S,(\tilde{x},x)\in R\Longrightarrow x\in X_S,\label{X_S}\\
&\Tilde{x}\in \Tilde{X}_F,(\tilde{x},x)\in R\Longrightarrow x\in X_F.\label{X_F}
\end{align}
\textcolor{black}{To synthesize a valid terminal control policy, we employ a safety game \cite{tabuada2009verification}, which is known to be an algorithmic technique to design a policy (for a finite transition system) such that the state trajectory remains in $\tilde{X}_F$ for all times.} The algorithm of the safety game is illustrated in \ralg{safety game(revise)}. 

\begin{algorithm}[t]
{
\caption{Terminal control policy synthesis for $\Tilde{T}_{\eta_S,\eta_I}$ (safety game).}\label{safety game(revise)}
\hspace*{0.02in}{\bf Input:}
$\Tilde{T}_{\eta_S,\eta_I},\Tilde{X}_F$\\
\hspace*{0.02in}{\bf Output:}
$\tilde{\pi}_{F}$
\begin{algorithmic}[1]
\State $n\leftarrow0,P^{(n)}\leftarrow \Tilde{X}_F$;
\While{$P^{(n+1)}\neq P^{(n)}$}
    
    \State $P^{(n+1)}\leftarrow P^{(n)}\cap{\rm Pre}_F(P^{(n)})$;
    \State $n\leftarrow n+1$; \\ 
{\bf end}
\EndWhile
\State $\Tilde{X}'_F\leftarrow P^{(n)}$; 
\For{$\tilde{x} \in \tilde{X}'_F$}
\State $\Tilde{\pi}_F (\Tilde{x})=\left\{(\tilde{u},\tilde{\varepsilon})\in \tilde{U}\times \tilde{E}\;|\; \forall \Tilde{x}'\in\Tilde{g}(\Tilde{x},\tilde{u},\tilde{\varepsilon}):\Tilde{x}'\in \tilde{X}'_F\right\}$;\\
{\bf end}
\EndFor 
\end{algorithmic}
}
\end{algorithm}

In the algorithm, the map ${\rm Pre}_F(P): 2^{\Tilde{X}_F}\rightarrow2^{\Tilde{X}_F}$ (line~3) is defined by 
\begin{equation}
\begin{gathered}
\label{pre_p}
{\rm Pre}_F(P)=\left\{\Tilde{x}=[\tilde{S},\tilde{I}]^{\top}\in \Tilde{X}_F\,|\,\exists \tilde{u}\in\tilde{U},\Tilde{\varepsilon}\in \tilde{E}:\right.\\
\phantom{=\;\;}\left. L_F(\Tilde{x},\Tilde{u},\Tilde{\varepsilon}) = 1, \forall{\Tilde{x}'\in \Tilde{g}(\Tilde{x},\tilde{u},\tilde{\varepsilon}),\Tilde{x}'\in P}\right\}.
\end{gathered}
\end{equation}
Hence, ${\rm Pre}_F(P)$ is the set of all states, for which there exists a pair of $\tilde{u}\in \tilde{U},\tilde{\varepsilon}\in \tilde{E}$, such that all the corresponding successors according to the transition function $\Tilde{g}$ are inside $P$ (i.e., $\Tilde{x}'\in P$) while fulfilling $L_F(\Tilde{x},\Tilde{u},\Tilde{\varepsilon}) = 1$. The condition $L_F(\Tilde{x},\Tilde{u},\Tilde{\varepsilon}) = 1$ is utilized to ensure that the controlled trajectory for not only the triggering time instants but also inter-event times is always inside $X_F$ 
(for details, see the proof of \rprop{terminal}). 
As shown in \ralg{safety game(revise)}, we iteratively compute $P^{(n)},n=0,1,...$, until the fixed point set is achieved (lines~2--5 in \ralg{safety game(revise)}). Note that this fixed point set is obtained within a finite number of iterations, since $\Tilde{X}_F, \tilde{U}$ are both finite. 
\textcolor{black}{In addition, since $\Tilde{g}(\Tilde{x},\tilde{u},\tilde{\varepsilon})$ is finite, 
we can construct $\tilde{\pi}_F (\tilde{x})$ (line~8) by collecting all pairs of $(\tilde{u}, \tilde{\varepsilon})$ such that $\Tilde{g}(\Tilde{x},\tilde{u},\tilde{\varepsilon}) \subseteq \tilde{X}'_F$ (i.e., every element in $\Tilde{g}(\Tilde{x},\tilde{u},\tilde{\varepsilon})$ belongs to $\tilde{X}_F$).}
Using $\tilde{\pi}_F$, we \textit{refine} the terminal control policies as follows. Let $X'_F$ be given by 
\begin{align}\label{X_T}
& X'_F=\left\{x\in X\;|\;\exists \tilde{x}\in\tilde{X}'_F:(\tilde{x},x)\in R\right\}. 
\end{align}
For any $x \in X'_F$, pick any $\tilde{x} \in {\rm Nearest}_{\Tilde{X}' _{F}}(x)$. Then, $\pi_F(x)$ is given as follows
\begin{align}\label{pi_T}
\pi_F(x)  = \{(u, \varepsilon)\ |\  u=\tilde{u},\varepsilon=\tilde{\varepsilon}, (\tilde{u},\tilde{\varepsilon})\in \Tilde{\pi}_F(\Tilde{x})\}. 
\end{align}

The following result shows that $\pi_F$ is a valid terminal event-triggered control policy. 

\begin{myprop}\label{terminal}
\normalfont
Suppose that \ralg{safety game(revise)} is implemented to obtain $X'_F \subseteq X_F$ and $\pi_F$. Then, $\pi_F$ is a valid terminal event-triggered control policy with respect to the set $X'_F$, i.e., $x(t_0)\in X'_F \implies \Phi^f_{\Pi_F,t}(x(t_0))\subseteq X'_F$ for all $t\in [t_0, \infty]$. \qedwhite 
\end{myprop}
The proof is given in the Appendix. 
\rprop{terminal} states that, once the state enters the set $X'_F$, it stays therein for all future times (which implies, since $X'_F \subseteq X_F$, it stays in $X_F$ for all future times). 
Next, we shall synthesize a valid \textit{reachable} event-triggered control policy, such that any controlled trajectory enters $X'_F$ in finite time while staying in $X_S$. To this end, we employ a reachability game \cite{tabuada2009verification}. 
The algorithm for the reachability game is illustrated in \ralg{reachability game(revise)}.
\begin{algorithm}[h]
{
\caption{Reachable control policy synthesis for $\Tilde{T}_{\eta_S,\eta_I}$ (reachability game).}\label{reachability game(revise)}
\hspace*{0.02in}{\bf Input:}
$\Tilde{T}_{\eta_S,\eta_I},\Tilde{X}_S,\Tilde{X}' _F$\\
\hspace*{0.02in}{\bf Output:}
$\tilde{\pi}_0, \tilde{\pi}$
\begin{algorithmic}[1]
\State $n\leftarrow0,\ Q^{(n)}\leftarrow \Tilde{X}_T$;
\While{$Q^{(n+1)}\neq Q^{(n)}$}
    \State $Q^{(n+1)}\leftarrow Q^{(n)}\cup{\rm Pre}(Q^{(n)})$;
    \State $\mathcal{N}(\tilde{x}) \leftarrow n+1$, $\forall \tilde{x} \in Q^{(n+1)} \backslash Q^{(n)}$; 
    \State $n\leftarrow n+1$; \\ 
{\bf end while}
\EndWhile
\State $\Tilde{X}' \leftarrow Q^{(n)}$;
\State $\Tilde{X}' _0\leftarrow{\rm Pre}_{0}(\Tilde{X}')$; 
\State $\mathcal{N}(\tilde{x}) \leftarrow n+1$, $\forall \tilde{x} \in \Tilde{X}'_0 \backslash \Tilde{X}'$; 
\For{$\tilde{x} \in \Tilde{X}'$}
\State $\Tilde{\pi} (\Tilde{x})=\left\{(\tilde{u},\tilde{\varepsilon})\in \tilde{U}\times \tilde{E}\;|\;\forall \Tilde{x}'\in\Tilde{g}(\Tilde{x},\tilde{u},\tilde{\varepsilon}): \mathcal{N}(\Tilde{x}') <  \mathcal{N}(\Tilde{x})\right\}$;\\
{\bf end}
\EndFor 
\For{$\tilde{x} \in \Tilde{X}'_0$}
\State $\Tilde{\pi}_0 (\Tilde{x})=\left\{(\tilde{u},\tilde{\varepsilon})\in \tilde{U}\times \tilde{E}\;|\;\forall \Tilde{x}'\in\Tilde{g}_0 (\Tilde{x},\tilde{u},\tilde{\varepsilon}):\mathcal{N}(\Tilde{x}') <  \mathcal{N}(\Tilde{x})\right\}$;\\ 
{\bf end}
\EndFor
\end{algorithmic}
}
\end{algorithm}
In the algorithm, the map ${\rm Pre}(Q): 2^{\Tilde{X}_S}\rightarrow2^{\Tilde{X}_S}$ (lines~3) is defined by 
\begin{equation}
\begin{gathered}
\label{pre_q}
{\rm Pre}(Q)=\left\{\Tilde{x}=[\tilde{S},\tilde{I}]^{\top}\in \Tilde{X}_S\,|\,\exists \tilde{u}\in\tilde{U},\Tilde{\varepsilon}\in \tilde{E}:\right.\\
\phantom{=\;\;}\left.\forall{\Tilde{x}'\in \Tilde{g}(\Tilde{x},\tilde{u},\tilde{\varepsilon}),\Tilde{x}'\in Q,\tilde{I}+\tilde{\varepsilon}\leq \bar{I}_S}\right\}
\end{gathered}
\end{equation}
Hence, ${\rm Pre}(Q)$ is the set of all states, for which there exists a pair of $\tilde{u}\in \tilde{U},\tilde{\varepsilon}\in \tilde{E}$, such that all the corresponding successors according to the transition function $\Tilde{g}$ are inside $Q$ (i.e., $\Tilde{x}'\in Q$) while fulfilling $\tilde{I}+\tilde{\varepsilon}\leq \bar{I}_S$. The condition $\tilde{I}+\tilde{\varepsilon}\leq \bar{I}_S$ is utilized to ensure that the controlled trajectory for not only the triggering time instants but also inter-event times is always inside $X_S$ 
(for details, see the proof of \rprop{reachable}). ${\rm Pre}_0 (\Tilde{X}')$ (line~8) is defined by 
\begin{equation}
\begin{gathered}
\label{pre_p0}
{\rm Pre}_0 (\Tilde{X}')=\left\{\Tilde{x}=[\tilde{S},\tilde{I}]^{\top}\in \Tilde{X}_S\,|\,\exists \tilde{u}\in\tilde{U},\Tilde{\varepsilon}\in \tilde{E}:\right.\\
\phantom{=\;\;}\left.\forall{\Tilde{x}'\in \Tilde{g}_0(\Tilde{x},\tilde{u},\tilde{\varepsilon}),\Tilde{x}'\in \Tilde{X}',\tilde{I}+\tilde{\varepsilon}\leq \bar{I}_S}\right\},
\end{gathered}
\end{equation}
We iteratively compute $Q^{(n)}$, $n=1, ...$ until the fixed point is achieved (lines~2--6 in \ralg{reachability game(revise)}). 
Now, let $X'_0, X'$ be given by 
\begin{align}\label{X'_0}
& X'_0=\left\{x\in X_0\;|\;\exists \tilde{x}\in\Tilde{X}':(\tilde{x},x)\in R_0 \right\},
\end{align}
\begin{align}\label{X_R}
& X'=\left\{x\in X\;|\;\exists \tilde{x}\in\Tilde{X}':(\tilde{x},x)\in R\right\}.
\end{align}
 For any $x\in X'_0$, pick any $\Tilde{x}\in {\rm Nearest}_{\Tilde{X}'_0}(x)$. 
Then, $\pi_0(x)$ is given as follows: 
\begin{align}\label{pi_R0}
& \pi_0(x)= 
\begin{cases}
\{(u,\varepsilon)\;|\;u=\tilde{u},\varepsilon=\tilde{\varepsilon}+(\tilde{I}-{I}),(\tilde{u},\tilde{\varepsilon})\in \Tilde{\pi}_0(\Tilde{x})\} \ (\mbox{if }L(\Tilde{x},\Tilde{u},\Tilde{\varepsilon}) = 1),\\
\{(u,\varepsilon)\;|\;u=\tilde{u},\varepsilon=\tilde{\varepsilon}+(I-\tilde{I}),(\tilde{u},\tilde{\varepsilon})\in \Tilde{\pi}_0(\Tilde{x})\}\ (\mbox{if }L(\Tilde{x},\Tilde{u},\Tilde{\varepsilon}) = 0).
\end{cases}
\end{align}
In addition, for any $x \in X'$, pick any $\tilde{x} \in {\rm Nearest}_{\Tilde{X}'}(x)$. Then, $\pi(x)$ is given as follows
\begin{align}\label{pi_R}
\pi(x)  = \{(u, \varepsilon)\ |\  u=\tilde{u},\varepsilon=\tilde{\varepsilon}, (\tilde{u},\tilde{\varepsilon})\in \Tilde{\pi}(\Tilde{x})\}. 
\end{align}
The following result shows that the pair of the above policies are proven to be valid.

\begin{myprop}\label{reachable}
\normalfont
Suppose that \ralg{reachability game(revise)} is implemented to obtain $\Pi =\{\pi_0, \pi\}$. Then, $\Pi$ is a valid pair of the reachable event-triggered control policies, if $X'_0 = X_0$. \qedwhite 
\end{myprop}
For the proof, see the Appendix. 
As the consequence, by designing valid terminal and the reachable event-triggered control policies based on \ralg{reachability game(revise)} and \ralg{safety game(revise)}, respectively, we can ensure that any controlled trajectory starting from any $x(t_0) \in X_0$ reaches $X'_F$ (or, $X_F$) in finite time (by applying $\Pi$), and then stays therein for all future times {(by applying $\pi_F$)}. 

\subsection{On the selection of the optimal pair in the policies}\label{selectionsec}
From the definition of the above control policies, there may exist multiple pairs of $({u},{\varepsilon})$ in $\pi$ (or $\pi_0$ as well as in $\pi_F$), and so one might wonder how to select the \textit{optimal} pair from $\pi$. A suitable way is to evaluate a cost with a look-ahead (including infinite) horizon, i.e., given the horizon length $T \in (0, \infty]$, 
\begin{align}
\underset{\{(u(t_\ell), \varepsilon(t_\ell))\}^{k+L} _{\ell=k}}{\mathrm{min}} & J_T (u(t_k), \ldots, u(t_{k+L})) \label{optfirst}\\ 
&\  \mathrm{s.t.}\  (u(t_\ell), \varepsilon(t_\ell)) \in \pi (x(t_\ell)) \\
&\ \ \ \ \ \ (u(t_\ell), \varepsilon(t_\ell)) \in \pi_0 (x(t_\ell))\ (\mathrm{if}\ \ell = 0) \\
&\ \ \ \ \ \ \ \ \ \ \ \ \forall \ell = k, \ldots, k+L, \label{optend}
\end{align}
where $L \in \mathbb{Z}_{>0}$ denotes the last index of the triggering time until the prediction horizon $t_k+T$ 
and the cost function
$J(u(t_k), \ldots, u(t_{k+L}))$ is given by 
\begin{align}
J(u(t_k), \ldots, u(t_{k+L})) &= \int^{t_k + T} _{t_k} \frac{\lambda^{t-t_k}}{u(t)} dt \notag  \\ 
&= \sum^L _{\ell=k+1} \left ( \frac{1}{u(t_{\ell-1})} \int^{t_{\ell}} _{t_{\ell-1}} {\lambda^{t-t_{k}}} dt \right ) + \frac{1}{u(t_L)} \int^{t_k +T} _{t_{L}} {\lambda^{t-t_k}} dt \notag \\
&= \sum^L _{\ell=k+1}\left ( \frac{\lambda^{\Delta_\ell} - \lambda^{\Delta_{\ell-1}}}{u(t_{\ell-1}) \ln \lambda} \right ) + \frac{\lambda^{T} - \lambda^{\Delta_L}}{u(t_L) \ln \lambda} \notag \\ 
&=\cfrac{1}{\ln \frac{1}{\lambda}}\left \{ \sum^L _{\ell=k+1}\left ( \frac{\lambda^{\Delta_{\ell-1}} - \lambda^{\Delta_{\ell}}}{u(t_{\ell-1})} \right ) + \frac{\lambda^{\Delta_L}-\lambda^T}{u(t_L)} \right \}, \label{cost2}
\end{align}
where $\Delta_\ell = t_\ell - t_k$ for all $\ell = k+1, ..., L-1$ and $\lambda \in (0, 1]$ is a decaying parameter so as to avoid the divergence of the cost to infinity if we set $T=\infty$. 
From the receding horizon perspective, after solving the above optimization problem only the optimal pair at the current time $t_k$, say $\{u^*(t_k), \varepsilon^*(t_k)\}$, is applied.

\begin{table}[tb]
\caption{\centering{Parameter Settings}}
\label{parameter}
\centering
\scalebox{0.9}{\begin{tabular}{c|c|c}
\hline
Parameter & Description & Value \\
\hline 
$u_0$ & no intervention & 0.26 \\
$u_1$ & pre-emergency & 0.22 \\
$u_2$ & state of emergency & 0.17 \\
$\gamma$ & recovery rate & 0.15 \\
$\xi$ & immunity loss rate & 0.02 \\
$\underline{S}_S$ & lower bound of $S$ for $X_S$ & 0.45 \\ 
$\overline{I}_S$ & upper bound of $I$ for $X_S$ & 0.10 \\
$\underline{S}_F$ & lower bound of $S$ for $X_F$ & 0.60 \\
$\overline{I}_F$ & upper bound of $I$ for $X_F$ & 0.05 \\
\end{tabular}}
\end{table}

\section{Example}\label{numerical simulation}
In this section, we evaluate the effectiveness of the proposed approach through a numerical simulation. The simulation is conducted in MATLAB R2022a running on Windows~10, Intel(R) Core(TM) i5-9400 CPU, 2.90GHz, 8GB RAM.
\subsection{Parameter setting}\label{numerical simulation parameter}
 The parameter settings are illustrated in \rtab{parameter}. As stated in \rsec{system description}, some previous studies have estimated the infection rate for each stage during the outbreak in Tokyo, Japan, obtaining a good fit with the infection data record. The application time period of each public measure is searchable from \cite{measure}. 
Based on these results, we provide the set of the control inputs as $u_0=0.26$ (no intervention), $u_1=0.22$ (pre-emergency), $u_2=0.17$ (state of emergency), the recovery rate is given by $\gamma=0.15$, {\color{black}and the immunity loss rate is given by $\xi=0.02$}. We construct the symbolic model $\Tilde{T}_{\eta_S,\eta_I}$ with the discretization parameters $\eta_{S}=\eta_{I}=0.01$. The set of the thresholds is defined as $\tilde{E}=\left\{0.01,0.02,0.03\right\} \cup \{0\}$. According to \cite{population}, the total population of Tokyo by January 1, 2023, is $N_{\rm total}=1.4\times 10^7$, and \cite{icu} shows that 1157 ICU spots are available in Tokyo, which we denote by $N_{\rm ICU}=1157$. The severity rate (i.e., the rate at which the infected subjects experience severe symptoms) is set as $\tau=0.08\%$. Hence, we set the upper bound of the fraction of infected subjects for $X_S$ as $\bar{I}_S=\frac{N_{\rm ICU}}{N_{\rm total}\times \tau} =0.10$. The other parameters such as $\underline{S}_S$, $\underline{S}_F$, $\bar{I}_F$ are shown in \rtab{parameter}\footnote{In general, the parameters $\underline{S}_S$, $\underline{S}_F$, $\bar{I}_F$ are chosen by the authorities after analyzing the intricate relationship between the social economy and the rate of susceptible (or infected) individuals. A more comprehensive method for their selection is beyond the scope of this paper. }.


\begin{figure*}[t]
    \centering
    \includegraphics[width=1.0\textwidth]{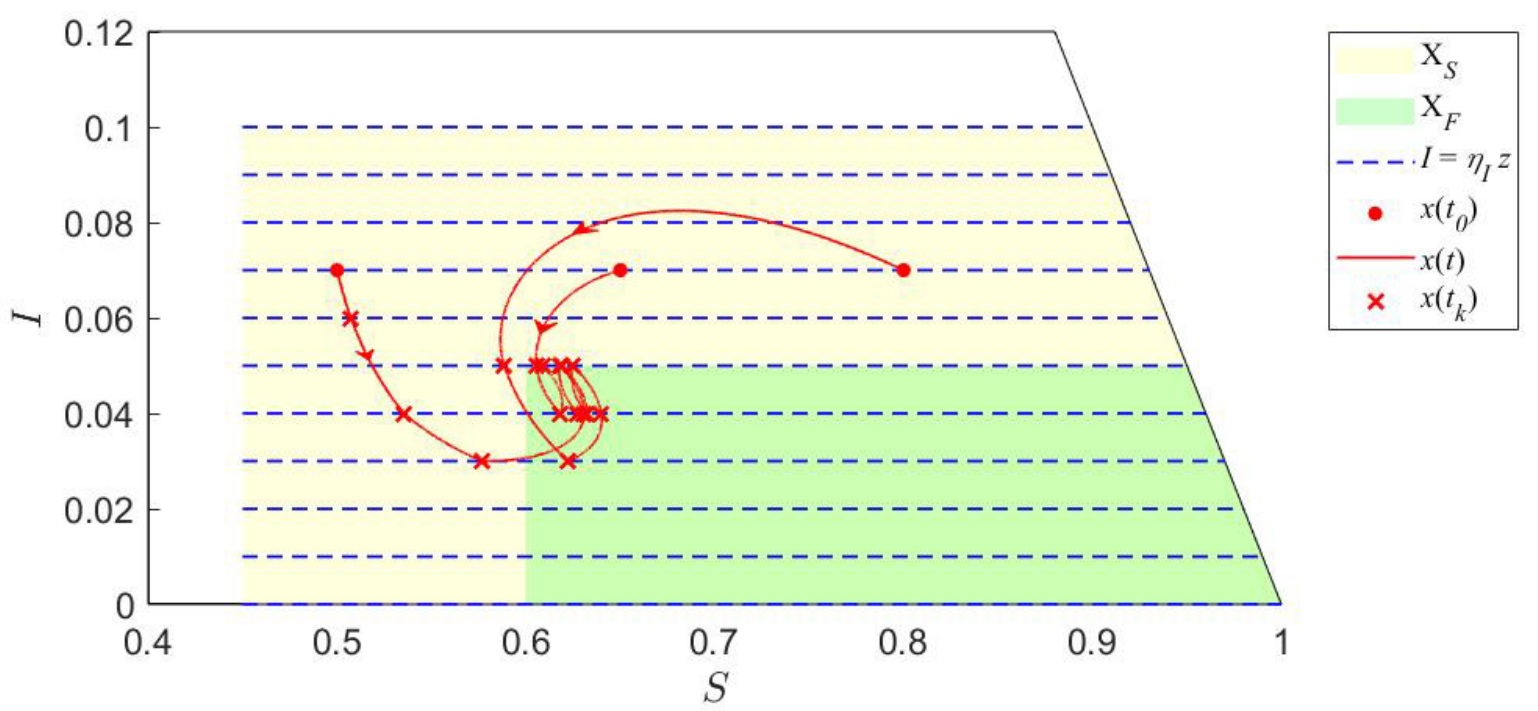}
    \caption{{Controlled trajectories by applying the proposed policy \req{optfirst}--\req{optend} (red solid lines), where the initial state is given by $[0.50,0.07]^\top$,$[0.65,0.07]^\top$ and $[0.80,0.07]^\top$.}}
    \label{yoko}
\end{figure*}

\begin{figure*}[t]
    \centering
    \includegraphics[width=1.0\textwidth]{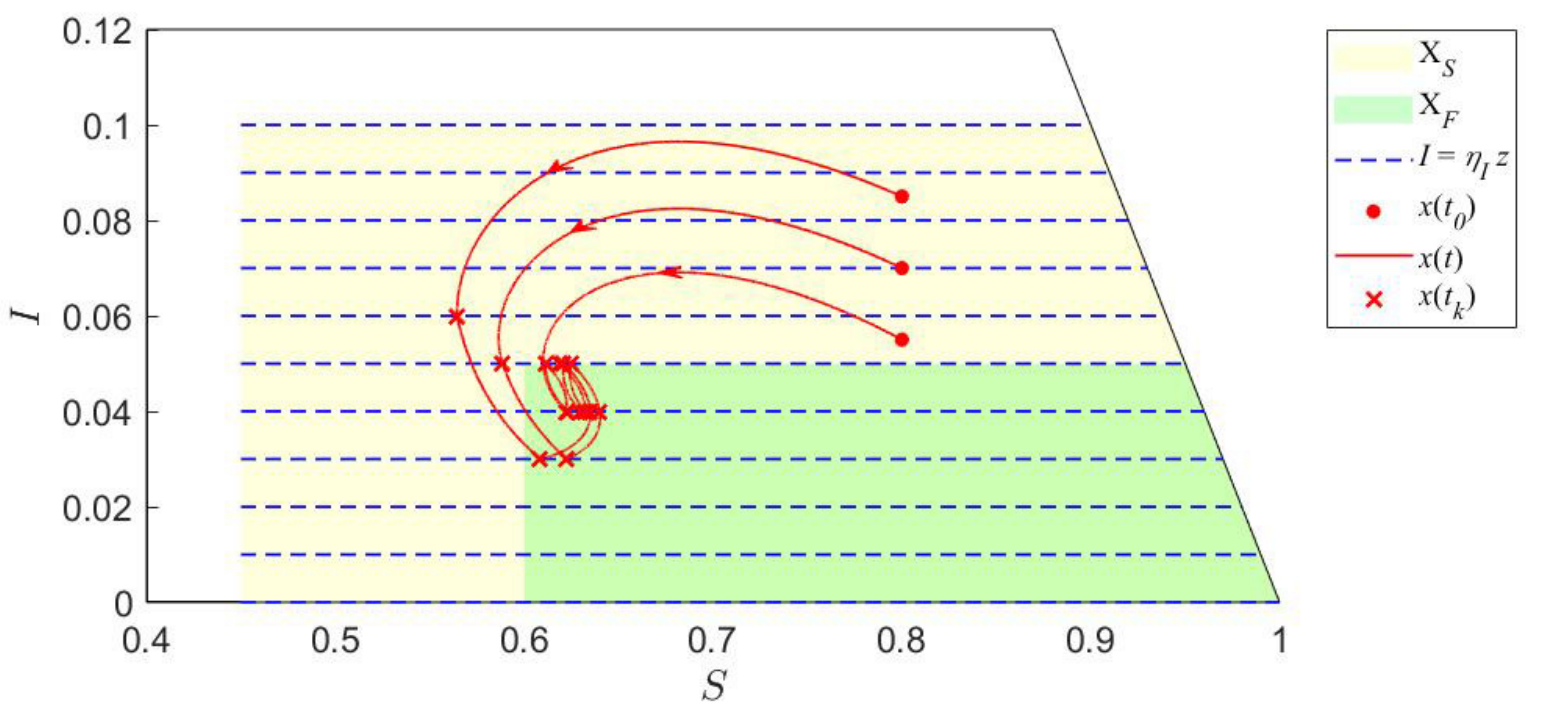}
    \caption{{State trajectories by applying the proposed policy \req{optfirst}--\req{optend} (red solid lines), where the initial state is given by $[0.80,0.055]^\top$,$[0.80,0.07]^\top$ and $[0.80,0.085]^\top$.}}
    \label{tate}
\end{figure*}

\subsection{Results and discussion} 
{\color{black}\rfig{yoko} and \rfig{tate} show the simulation results. 
The pale yellow area illustrates $X_S$ given by \req{xsdef}. The cyan area illustrates $X_F$ given by \req{xfdef}. The blue dashed lines illustrate the ones for which the event can trigger (i.e., $I = \eta_I z$, where $z$ denotes some non-negative integer).}
\begin{figure*}
    \centering
    \subfigure[{Controlled trajectories in the $S-I$ plane}]{
    \begin{minipage}{14cm}
    \centering
        \includegraphics[width=1.0\textwidth]{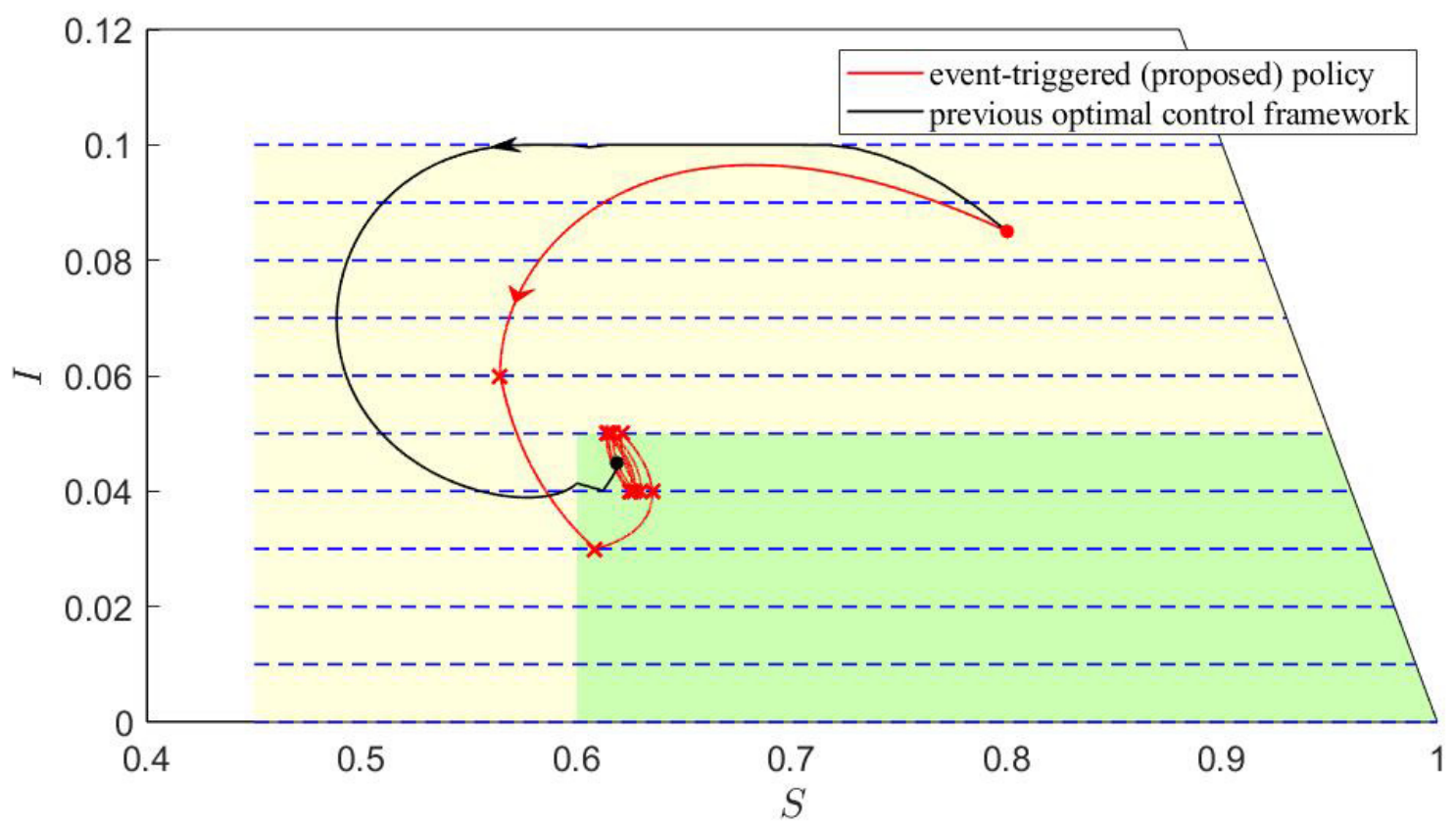}
        \end{minipage}
    }

    \subfigure[{Fraction of susceptible/infected subjects}]{
    \begin{minipage}{14cm}
    \centering
        \includegraphics[width=1.0\textwidth]{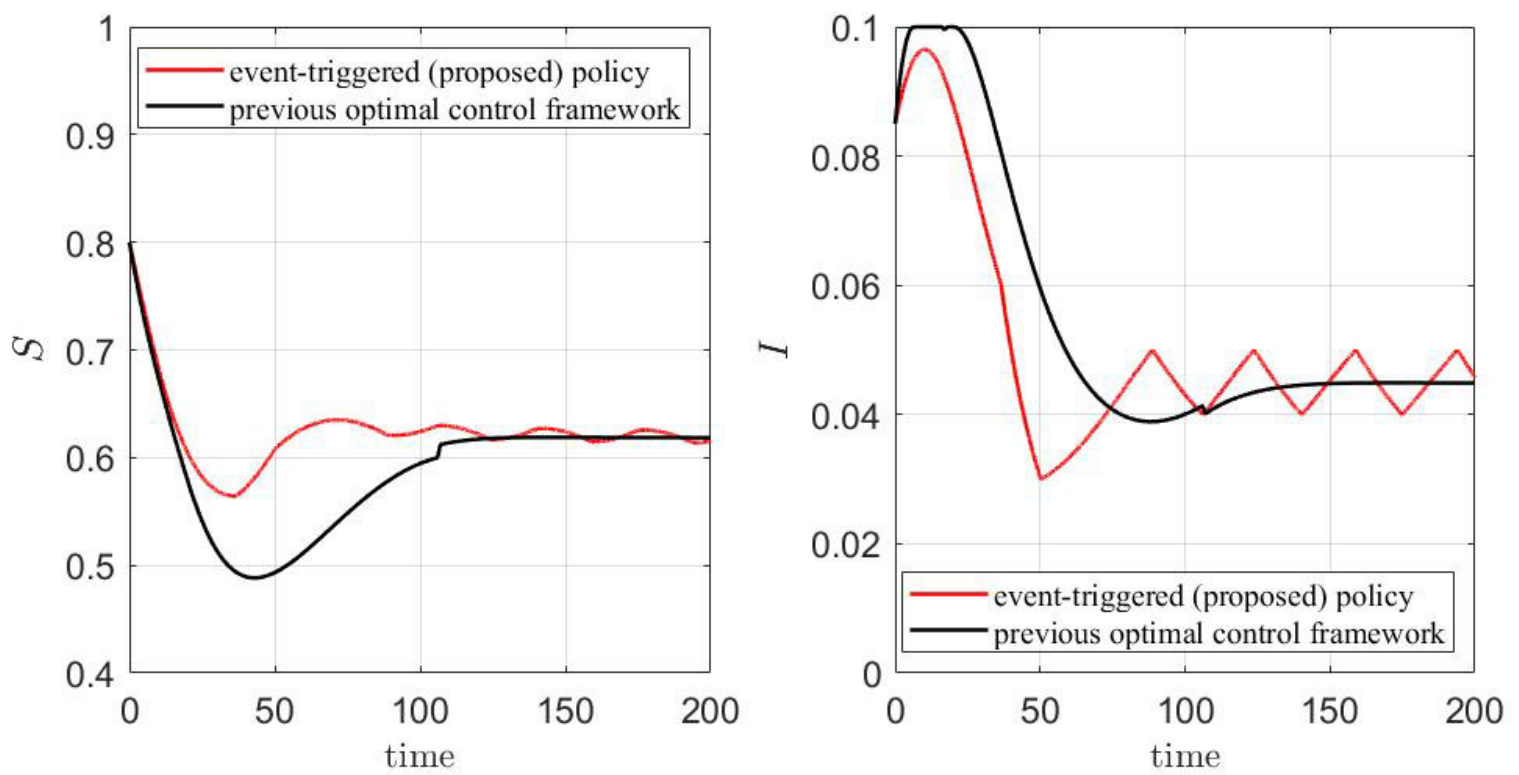}
        \end{minipage}
    }
    \caption{{(a) State trajectories by applying the proposed policy \req{optfirst}--\req{optend} (red solid line) and previous optimal control framework (black solid line) with the initial state $x_0 = [0.80, 0.07]^\top$. The arrows indicate the direction of the trajectories, and the red cross marks represent the triggering instants. 
    (b) State trajectories of the fraction of susceptible/infected subjects against time $t$.}}
    \label{compare}
\end{figure*}

\begin{figure*}[t]
    \centering
    \subfigure[{Applied control inputs}]{
    \begin{minipage}{12cm}
        \includegraphics[width=1.0\textwidth]{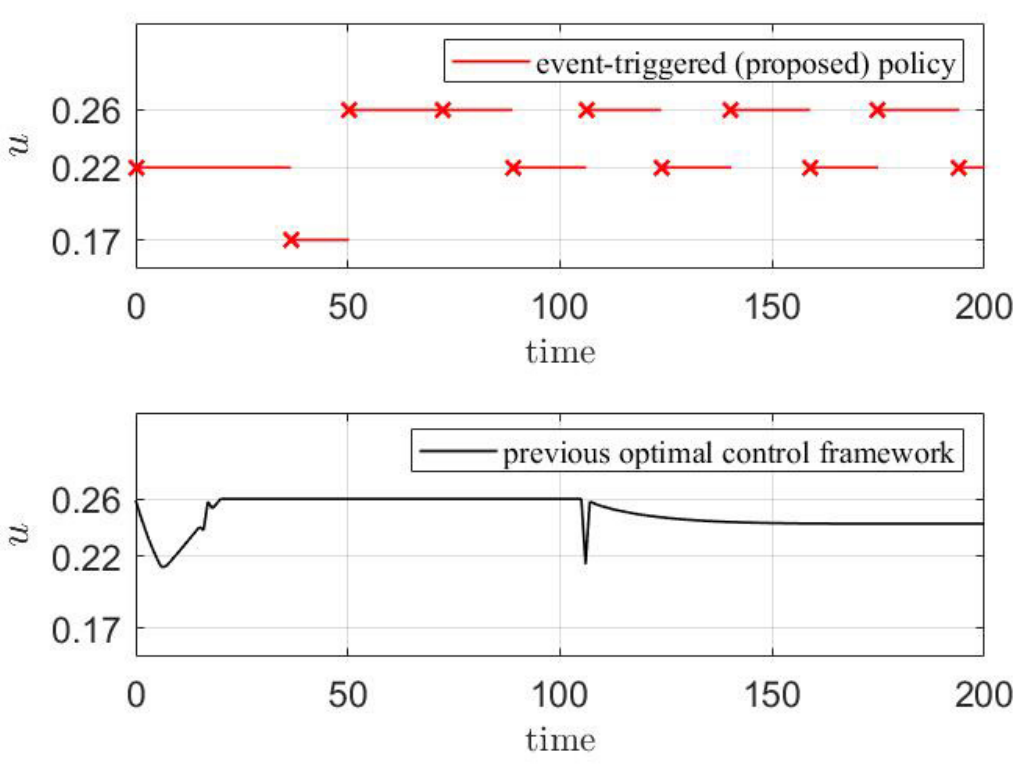}
    \end{minipage}
    }
    
    \caption{{Control input by applying the proposed approach \req{optfirst}--\req{optend} (upper figure) and previous optimal control (lower figure).}}
    \label{cost}
\end{figure*}
{\color{black}During control execution, the pair of the control input and the threshold is chosen from the obtained policies according to the proposed policy \req{optfirst}--\req{optend} with $\lambda = 0.99$ and $T= \infty$. The red solid lines in \rfig{yoko} represent some controlled trajectories with the initial state 
$x(t_0)$ being given by $[0.50,0.07]^\top$,$[0.65,0.07]^\top$,$[0.80,0.07]^\top$. The red solid lines in \rfig{tate} represent some controlled trajectories with the initial state $x(t_0)$ being given by $[0.80,0.055]^\top$,$[0.80,0.07]^\top$,$[0.80,0.085]^\top$.
In addition, the red cross points denote the states at the triggering time instants. 
It has been shown from the figure that all the controlled trajectories enter the set $X_F$ in finite time and stay therein for all future times.}

{\color{black} We compare the proposed approach with the previous optimal control-based framework (see, e.g., \cite{gaff2009}), where the control input is presumed to be \textit{continuously valued and updated}.
Specifically, the optimal control trajectory is obtained by solving the optimization problem: 
\begin{align}\label{previous}
    \underset{u(\cdot) \in U_c}{\mathrm{min}} \int^{t + T} _{0} \frac{\lambda^{\tau}}{u(\tau)} d\tau 
\end{align}
subject to the dynamics \req{system_f} and $x(T) \in X_F$, $x(t) \in X_S$ for all $t \in [0, T]$, and the control input is restricted within the {continuous} range $U_c = [0.17,0.26]$. 
The decaying parameter is set as $\lambda=0.99$, and the horizon length is set as $T=200$. 
Note that the necessary condition of optimality can be obtained by solving the corresponding Euler-Lagrange equation under the Pontryagin maximum principle (see, e.g., \cite{gaff2009}). 
We compare two controlled trajectories in the upper figure of \rfig{compare}, where the red solid trajectory is generated according to \req{optfirst}--\req{optend}, and the black solid trajectory is generated by the previous optimal control method, with the same initial state $x_0 = [0.80, 0.085]^\top$. 
\rfig{cost} shows the control inputs applied according to \req{optfirst}--\req{optend} (red solid lines) and the previous optimal control method (black solid line). 
For the previous optimal control method, ``no intervention" (i.e., $u(t) = 0.26$) is recommended constantly for some time and thereafter it ends up with an equilibrium point (black dot in the upper figure of \rfig{compare}). The applied control input converges to some constant value within the range $(0.22, 0.26)$, i.e., $\lim_{t\rightarrow \infty}u(t) \in (0.22, 0.26)$. From \rfig{cost}, the previous optimal control method provides a better control performance than our approach in terms of the cost \req{previous}, in the sense that it tends to apply larger control inputs (e.g., ``no intervention'') than our approach. This may be due to the fact that our approach requires the discretization of the state space to design an event-triggered control policy (which may lead to a more conservative policy than the previous methods) and that the previous optimal control method is allowed to use the continuous-valued control input. 
Nevertheless, our approach is advantageous over the previous optimal control method, in the sense that it allows the control input to take \textit{discrete} values and to be updated \textit{only when} the fraction of the infected people changes the designed threshold (according to our event-triggered strategy), which is more natural and appropriate for mitigating the epidemics.}

\section{Conclusion}\label{conclusion and future work}
This paper proposed an event-triggered strategy for the SIRS epidemic model, handling the problem of epidemic control with manageable public measures. In the proposed strategy, control inputs that indicate the level of the public measure are updated only when the fraction of infected subjects increases or decreases by a prescribed threshold. \textcolor{black}{In particular, we set an upper (resp. lower) bound against the fraction of infected (resp. susceptible) subjects, and we aim to control the fraction of infected (susceptible) subjects within a specific level in finite time.} To this aim, we first defined an event-triggered transition system to capture the dynamical behavior of the susceptible, infected, and removed subjects with respect to the event-triggered strategy, based on which we constructed the symbolic model. The event-triggered policies were synthesized via the safety and the reachability games, and we derived a suitable condition for the event-triggered policies to be valid. 
\textcolor{black}{Finally, the effectiveness of the event-triggered control approach was verified through a numerical simulation.}

\section*{Acknowledgement}
This work is supported by JST CREST JPMJCR201, Japan and by JSPS KAKENHI Grant 21K14184.


\appendix
\section{Proof of Proposition~3}
Assume $x(t_0) =[S(t_0), I(t_0)]^\top 
 \in X'_F$. Then, there exists $\Tilde{x}(t_0) = [\tilde{S}(t_0), \tilde{I}(t_0)]^\top \in \tilde{X}'_F$ such that $(\tilde{x}(t_0), x(t_0)) \in R$. 
Note that we have $I(t_0) = \tilde{I}(t_0)$ and $|S(t_0)-\tilde{S}(t_0)|\leq \eta_S/2$. 
Pick $\left\{\tilde{u}(t_0),\tilde{\varepsilon}(t_0)\right\}\in \Tilde{\pi}_{F} (\Tilde{x}(t_0))$ and $\left\{u(t_0),\varepsilon(t_0)\right\}\in \Tilde{\pi}_{F} (x(t_0))$ satisfying $u(t_0)=\tilde{u}(t_0)$, $\varepsilon(t_0)=\tilde{\varepsilon}(t_0)$. Then, consider applying the event-triggered strategy, i.e.,  $u_0$ is applied until the next triggering time instant 
\begin{align}
t_1 = \mathrm{inf} \left\{t>t_k:|I(t)-I(t_0)| \geq \varepsilon(t_0) \right\}, 
\end{align}
i.e., the time instant when the fraction of infected subjects changes over $\varepsilon(t_0)$. 
Pick $\tilde{x}(t_1) = [\tilde{S}(t_1), \tilde{I}(t_1)]^\top$ satisfying $(\tilde{x}(t_1), x(t_1)) \in R$. Since $(R_0, R)$ is $(\eta_S, \eta_I)$-ASR and the pair $\left\{\tilde{u}(t_0),\tilde{\varepsilon}(t_0)\right\}$ is chosen from the control policy $\Tilde{\pi}_{F} (\Tilde{x}(t_0))$ according to line~8 in \ralg{safety game(revise)}, it follows that $\tilde{x}(t_1) \in \tilde{g}_0 (\tilde{x}(t_0), \tilde{u}(t_0), \tilde{\varepsilon}(t_0))$. 
Regarding the next triggering state $x(t_1)$, we have the following two cases: (a) $I(t_1) = I(t_0) + \varepsilon(t_0)$ ($I(t)$ increases by $\varepsilon(t_0)$); (b) $I(t_1) = I(t_0) - \varepsilon(t_0) >0$ ($I(t)$ decreases by $\varepsilon(t_0)$). 
For case~(a), ${x}(t_0) \in {X}'_F$ and ${x}(t_1) \in {X}'_F$ implies $I(t_0) \leq \bar{I}_F$, $I(t_1) = I(t_0) + \varepsilon(t_0) \leq \bar{I}_F$ since $X'_F\subseteq X_F$ (see the definition of $X_F$ in \req{xfdef}). 
Since $t_1$ is chosen to be the \textit{first} time instant when the fraction of infected subjects changes over $\varepsilon(t_0)$, it follows that $I(t_0) \leq \bar{I}_F$, $I(t_1) \leq \bar{I}_F \implies I(t) \leq \bar{I}_F$ for all $t \in [t_0, t_1]$. 
\textcolor{black}{In addition, from \req{pre_p} we have $L_F(\Tilde{x},\Tilde{u},\Tilde{\varepsilon}) = 1$, it follows that $\underline{S}_F \leq S(t)$ for all $t \in [t_0, t_1]$.}
Therefore, we obtain $x(t) = [S(t), I(t)]^\top \in X_F$ for all $t \in [t_0, t_1]$.
For case~(b), we have $I(t_0) \leq \bar{I}_F$ and $I(t_1) = I(t_0) - \varepsilon(t_0) \leq \bar{I}_F$. 
Since $I(t_0)=\tilde{I}(t_0),\varepsilon(t_0)=\tilde{\varepsilon}(t_0)$, based on condition~(ii) in \rdef{symbolic model}, we have $I(t) < \tilde{I}(t_0) + \tilde{\varepsilon}(t_0) = I(t_0) + \varepsilon(t_0)$ for all $t \in [t_0, t_1]$. In addition, from \req{pre_p}, we have $I(t_0)+\varepsilon(t_0)\leq \bar{I}_F$ and thus $I(t) < \bar{I}_F$ for all $t \in [t_0, t_1]$. \textcolor{black}{As with case~(a), we have $\underline{S}_F \leq S(t)$ for all $t \in [t_0, t_1]$.} Therefore, we obtain $x(t) = [S(t), I(t)]^\top \in X_F$ for all $t \in [t_0, t_1]$. Thus, we obtain $x(t) = [S(t), I(t)]^\top \in X_F$ for all $t \in [t_0, t_1]$. 
By recursively applying exactly the same procedure as above, we have $\tilde{x}(t_0)\in \tilde{X}' _F \implies  \tilde{x}(t_1) \in \tilde{X}' _F$, $\tilde{x}(t_2) \in \tilde{X}'_F,...$, and then $x(t_0) \in X'_F \subseteq X_F \implies x(t) \in X'_F \subseteq X_F$ for all $t \geq t_0$. Hence, $\pi_F$ is a valid terminal event-triggered control policy if $x(t_0) \in X' _F$.   \qedwhite

\section{Proof of Proposition~4}
Assume $X_0 = X'_0$ and let $x(t_0) =[S(t_0), I(t_0)]^\top \in X_0$ so that we have $x(t_0)\in X'_0$. Then, there exists $\Tilde{x}(t_0) = [\tilde{S}(t_0), \tilde{I}(t_0)]^\top \in \tilde{X}'_0$ such that $(\tilde{x}(t_0), x(t_0)) \in R_0$. 
Note that $|I(t_0) - \tilde{I}(t_0)| \leq \eta_I/2$ and $|S(t_0)-\tilde{S}(t_0)|\leq \eta_S/2$. 
Pick $\left\{\tilde{u}(t_0),\tilde{\varepsilon}(t_0)\right\}\in \Tilde{\pi}_{0} (\Tilde{x}(t_0))$ and $\left\{u(t_0),\varepsilon(t_0)\right\}\in \pi_{0} (x(t_0))$ satisfying $u(t_0)=\tilde{u}(t_0)$, $\varepsilon(t_0)=\tilde{\varepsilon}(t_0)+\tilde{I}(t_0) - I(t_0)$ (if $L(\Tilde{x},\Tilde{u},\Tilde{\varepsilon}) = 1$) or $\varepsilon(t_0)=\tilde{\varepsilon}(t_0)-I(t_0)-\tilde{I}(t_0)$ (if $L(\Tilde{x},\Tilde{u},\Tilde{\varepsilon}) = 0$). 
Then, consider applying the event-triggered strategy, i.e.,  $u_0$ is applied until the next triggering time instant 
\begin{align}
t_1 = \mathrm{inf} \left\{t>t_k:|I(t)-I(t_0)| \geq \varepsilon(t_0) \right\}, 
\end{align}
i.e., the time instant when the fraction of infected subjects changes over $\varepsilon(t_0)$. 
Pick $\tilde{x}(t_1) = [\tilde{S}(t_1), \tilde{I}(t_1)]^\top$ satisfying $(\tilde{x}(t_1), x(t_1)) \in R$. Since $(R_0, R)$ is $(\eta_S, \eta_I)$-ASR and the pair $\left\{\tilde{u}(t_0),\tilde{\varepsilon}(t_0)\right\}$ is chosen from the control policy $\Tilde{\pi}_{0} (\Tilde{x}(t_0))$ according to line~14 in \ralg{reachability game(revise)}, it follows that $\tilde{x}(t_1) \in \tilde{g}_0 (\tilde{x}(t_0), \tilde{u}(t_0), \tilde{\varepsilon}(t_0))$ and  there exists $n_0 \leq n$ such that $\mathcal{N} (\tilde{x}(t_1)) = n_0$, i.e., $\tilde{x}(t_1) \in Q^{(n_0)} \backslash Q^{(n_0-1)} \subseteq \mathrm{Pre}(Q^{(n_0-1)})$. 
Regarding the next triggering state $x(t_1)$, we have the following two cases: (a) $I(t_1) = I(t_0) + \varepsilon(t_0)$ ($I(t)$ increases by $\varepsilon(t_0)$); (b) $I(t_1) = I(t_0) - \varepsilon(t_0) >0$ ($I(t)$ decreases by $\varepsilon(t_0)$). 
For case~(a), ${x}(t_0) \in {X}^0_R$ and ${x}(t_1) \in {X}_R$ implies $I(t_0) \leq \bar{I}_S$, $I(t_1) = I(t_0) + \varepsilon(t_0) \leq \bar{I}_S$ since $X'_0\subseteq X_S, X_R\subseteq X_S$ (see the definition of $X_S$ in \req{xsdef}). 
Since $t_1$ is chosen to be the \textit{first} time instant when the fraction of infected subjects changes over $\varepsilon(t_0)$, it follows that $I(t_0) \leq \bar{I}_S$, $I(t_1) \leq \bar{I}_S \implies I(t) \leq \bar{I}_S$ for all $t \in [t_0, t_1]$. 
In addition, from condition~(i) in \rdef{symbolic model},
it follows that $\underline{S}_S \leq S(t)$ for all $t \in [t_0, t_1]$. 
Therefore, we obtain $x(t) = [S(t), I(t)]^\top \in X_S$ for all $t \in [t_0, t_1]$. 
For case~(b), we have $I(t_0) \leq \bar{I}_S$ and $I(t_1) = I(t_0) - \varepsilon(t_0) \leq \bar{I}_S$. 
Since $I(t_0)=\tilde{I}(t_0),\varepsilon(t_0)=\tilde{\varepsilon}(t_0)$, based on condition~(ii) in \rdef{symbolic model}, we have $I(t) < \tilde{I}(t_0) + \tilde{\varepsilon}(t_0) = I(t_0) + \varepsilon(t_0)$ for all $t \in [t_0, t_1]$. In addition, from \req{pre_p0}, we have $I(t_0)+\varepsilon(t_0)\leq \bar{I}_S$ and thus $I(t) < \bar{I}_S$ for all $t \in [t_0, t_1]$. Moreover, based on condition~(ii) in \rdef{symbolic model}, we have $\underline{S}_S \leq S(t)$ for all $t \in [t_0, t_1]$. Therefore, we obtain $x(t) = [S(t), I(t)]^\top \in X_S$ for all $t \in [t_0, t_1]$. Thus, we obtain $x(t) = [S(t), I(t)]^\top \in X_S$ for all $t \in [t_0, t_1]$. 
Now, there exists $\Tilde{x}(t_1) = [\tilde{S}(t_1), \tilde{I}(t_1)]^\top \in \Tilde{X}'$ such that $(\tilde{x}(t_1), x(t_1)) \in R$. Pick $\left\{\tilde{u}(t_1),\tilde{\varepsilon}(t_1)\right\}\in \Tilde{\pi}_R (\Tilde{x}(t_1))$ and $\left\{u(t_1),\varepsilon(t_1)\right\}\in \pi_R (x(t_1))$ satisfying $u(t_1)=\tilde{u}(t_1)$, $\varepsilon(t_1)=\tilde{\varepsilon}(t_1)$, and apply the event-triggered control policy to obtain $x(t_2)$. Pick $\tilde{x}(t_2) = [\tilde{S}(t_2), \tilde{I}(t_2)]^\top$ satisfying $(\tilde{x}(t_2), x(t_2)) \in R$.
As with the case for $t_1$, we have $\tilde{x}(t_2) \in \tilde{g}(\tilde{x}(t_1), \tilde{u}(t_1), \tilde{\varepsilon}(t_1))$. Moreover, since $\tilde{x}(t_1) \in Q^{(n_0)} \backslash Q^{(n_0-1)} \subseteq \mathrm{Pre}(Q^{(n_0-1)})$, we have $\tilde{x}(t_2) \in Q^{(n_0-1)}$. By employing the same procedure with the case $t_0$, it is shown that $x(t) \in X_S$ for all $t \in [t_1, t_2]$. 
Therefore, by recursively applying the same procedure as above, we have $\tilde{x}(t_0)\in \Tilde{X}' \implies  \tilde{x}(t_1) \in Q^{(n_0)}$, $\tilde{x}(t_2) \in Q^{(n_0-1)}$, $\tilde{x}(t_3) \in Q^{(n_0-2)},...,\tilde{x}(t_{n_0+1}) \in Q^{(0)} = \tilde{X}_T(\subseteq \tilde{X}_F)$, and the state trajectory by applying the event-triggered control policy satisfies $(\tilde{x}(t_k), x(t_k)) \in R$ for all $k=1, 2, ...$ and thus $x(t_{n_0+1}) \in X_T(\subseteq X_F)$. In addition, it is shown that $x(t) \in X_S$ for all $t \in [t_0, t_{n_0+1}]$. 
Hence, $\Pi_ = \{\pi_0, \pi\}$ is a valid pair of the reachable event-triggered control policies.  \qedwhite





\end{document}